# Least ln-cosh–based adaptive filtering algorithm for graph signal processing

**Chong Zhang**

The Key Laboratory of Magnetic Suspension Technology and Maglev Vehicle, Ministry of Education, School of Electrical Engineering, Southwest Jiaotong University, Chengdu, China.

**Haiquan Zhao, Senior Member, IEEE**

The Key Laboratory of Magnetic Suspension Technology and Maglev Vehicle, Ministry of Education, School of Electrical Engineering, Southwest Jiaotong University, Chengdu, China.

***Abstract*—Currently, an increasing number of irregular signals are being defined within graph structures. This paper proposes a Least ln-cosh-based Adaptive Filtering for Graph Signal Processing (GSP LLAF) algorithm for offline/online graph signal estimation under various non-Gaussian noise conditions. Compared to the recently proposed Graph Adaptive Least Mean p-th Power (GSP LMP) algorithm, GSP LLAF eliminates the need for prior knowledge of the noise characteristics. Simultaneously, GSP LLAF addresses the limitation in estimation accuracy encountered by the Sign Algorithm for Graph Signal Processing (GSP Sign) and the Graph Signal Processing algorithm based on the Generalized Maximum Correntropy Criterion (GSP GMCC). Furthermore, the algorithm features low computational complexity and fast execution speed. To achieve a balance between convergence speed and estimation accuracy, a variable step-size variant, the Modified Variable Step-Size Least ln-cosh-based Adaptive Filtering Algorithm for Graph Signal Processing (GSP MVSS-LLAF), is proposed. In estimation experiments involving both static and time-varying graph signals, the proposed GSP LLAF and GSP MVSS-LLAF algorithms demonstrate fast, stable, and robust graph signal estimation performance under various non-Gaussian noise conditions, including Bernoulli-Gaussian (BG) modeled impulse noise, as well as alpha-stable, Cauchy, and Laplace distributions.**

## I. INTRODUCTION

Signal processing finds wide application in communications [1], [2], image and audio analysis [3]. Modern irregular-domain data are no longer confined to simple one- or two-dimensional spaces; examples include spatial temperature measurements, sensor networks, power-system data, computer vision, and biomedical applications [4], [5], [6], [7], [8], [9]. Increasingly, irregular signals are defined on graphs. Graph Signal Processing (GSP) extends digital signal processing concepts to graphs and has been applied across diverse domains to process irregular data, including filtering, distributed semi-supervised learning, classification, graph-filter design, and adaptive graph-signal estimation [10], [11], [12], [13], [14]. Because signals are corrupted by noise during acquisition and transmission, many graph signals require online estimation; accordingly, Adaptive Filtering (AF) concepts [15], [16], [17], [18] have been combined with GSP to enable online estimation and tracking of static or time-varying graph signals [14]. Moreover, measurement and transmission noise is often non-Gaussian, with common models including alpha-stable, Laplace, Cauchy, and Bernoulli-Gaussian (BG) modeled impulsive noise [19], [20], [21], [22], [23]. Integrating AF with robust statistical criteria [24], [25], [26] in the graph domain not only mitigates the limitations of conventional methods under irregular topologies and non-Gaussian noise, but also substantially improves the robustness and real-time responsiveness of practical systems in such environments.

Least Mean Square (LMS) and normalized least mean square (NLMS) strategies have been introduced into the graph-frequency domain, yielding the GSP LMS and GSP NLMS algorithms and establishing a framework that integrates the LMS paradigm with the Graph Fourier Transform (GFT) representation, which serves as the baseline for subsequent graph-adaptive methods [4], [14]. Although GSP LMS and GSP NLMS perform well under Gaussian noise, their performance degrades severely or even diverges in non-Gaussian environments, and both methods rely strongly on the spectral-sparsity assumption. To address these limitations, the Generalized Maximum Correntropy Criterion (GMCC) [27] was mapped to the graph-frequency domain to produce GSP GMCC, and further augmented with Variable Step Size (VSS) [28] and Variable Kernel Width (VKW) [29] strategies to form the GSP VSS–VKW GMCC algorithm [30]. The GMCC function itself is non-convex and non-concave, involves multiple parameters, and presents complex numerical solutions. It is prone to falling into local optima and may result in slow transient responses. Therefore, it entails a trade-off between robustness and convergence speed. For α-stable and other heavy-tailed noise, the adaptive estimation problem on graphs has been cast under a Minimum Dispersion (MD) criterion, yielding the Least Mean *p*-th power for GSP (GSP LMP) algorithm; LMP enables a tunable trade-off between robustness and computational complexity but requires computing *p*-th powers and its optimal *p* typically depends on noise priors, often requiring manual tuning in practice [31]. To overcome LMP's limitations, the Sign for GSP (GSP Sign) algorithm was proposed: by replacing error magnitudes with their signs the method effectively clips large outliers and reduces computational cost, and its normalized variant Normalized Sign for GSP (GSP NSign) improves convergence speed [32], [33]. However, the sign function's non-smoothness and the loss of magnitude information can

Manuscript received XXXXX 00, 0000; revised XXXXX 00, 0000; accepted XXXXX 00, 0000.

This work was partially supported by the National Natural Science Foundation of China (grant: 62671522, 62171388, 61871461, 61571374) and Supported by the Joint Innovat-ion Fund Project ofLanzhou Jiaotong University and Corresponding Supporting University (grant: 2682026ZTZ002) (*Corresponding author*: Haiquan Zhao).

Chong Zhang (e-mail: chongzhang12@126.com) and Haiquan Zhao (e-mail: hqzhao_swjtu@126.com) are with the Key Laboratory of Magnetic Suspension Technology and Maglev Vehicle, Ministry of Education, School degrade steady-state accuracy in low-impulsive-noise regimes and limit effectiveness against other non-Gaussian noise types. Moreover, graph signal recovery has also been investigated through alternative paradigms, such as gain-weighted adaptive estimation and probabilistic inference [34], [35]. The former requires the additional design and updating of proportionate gains, and its performance may depend on signal sparsity, whereas the latter typically relies on prespecified probabilistic models and iterative inference, making these approaches more complex than robust loss-driven graph adaptive filtering. To address the above issues, this paper applies a least ln-cosh adaptive filtering (LLAF) cost function to the field of graph signal processing. The ln-cosh criterion provides a smooth and convex compromise between MSE and MAE estimation. Its gradient, $\tanh(pe)$, preserves error-magnitude information for small residuals while smoothly limiting the influence of large outliers. Therefore, it avoids the non-smooth updates of the Sign criterion and the non-convex optimization associated with correntropy-based criteria.

This paper makes the following contributions:

1)We derive the GSP LLAF algorithm for offline and online recovery of bandlimited graph signals from partial and noisy observations, providing a smooth and bounded-influence update under non-Gaussian noise.

2)We propose a modified variable step-size least ln-cosh-based adaptive filtering algorithm for graph signal processing (GSP MVSS-LLAF), as an extension for static scenarios with sufficient sampling set information.

3)Theoretical analyses related to the GSP LLAF and GSP MVSS-LLAF algorithms are conducted, establishing the step size constraint range and steady-state behavior of the GSP LLAF algorithm, its computational complexity, and the steady-state behavior of the update term for the GSP MVSS-LLAF algorithm. The superior performance of the algorithms under non-Gaussian noise is verified through experiments on static and time-varying graph signal estimation.

The organization of the paper is as follows. Section II details the fundamental knowledge of Graph Signal Processing. The proposal and derivation of the GSP LLAF algorithm and the proposal of the GSP MVSS-LLAF algorithm will be presented in Section III. Section IV provides the related theoretical analysis for both GSP LLAF and GSP MVSS-LLAF. Section V presents computer simulations. Finally, Section VI concludes the paper.

## II. Preliminaries on Graph Signal Processing

Consider a graph $\mathbf{G}$ with N nodes. The node signal defined on $\mathbf{G}$ is referred to as the graph signal $\mathbf{x} \in \mathbf{R}^{N\times 1}$. Its adjacency matrix $\mathbf{A} \in \mathbf{R}^{N\times N}$ is composed of the weights between nodes, representing the structure of the graph. The element $a_{mn}$ in the m-th row and n-th column of matrix $\mathbf{A}$ represents the connection strength between the m-th and n-th nodes. The graph Fourier transform (GFT) of $\mathbf{x}$ is based on the eigendecomposition of the adjacency matrix. Decompose $\mathbf{A}$ into

$$\mathbf{A} = \mathbf{U}\Lambda\mathbf{U}^T \tag{1}$$

where $\mathbf{U} = [\mathbf{u}_1, \mathbf{u}_2, \mathbf{u}_3, ..., \mathbf{u}_N]$ is the orthogonal eigenvector matrix and $\mathbf{\Lambda} = \text{diag}\{\lambda_1, \lambda_2, ..., \lambda_N\}$ is the diagonal eigenvalue matrix. The Graph Fourier Transform (GFT) [36], [37] is defined as

$$\mathbf{s} = \mathbf{U}^T\mathbf{x} \tag{2}$$

where $\mathbf{s} \in \mathbf{R}^{N\times 1}$ is the representation in the frequency domain. The Inverse Graph Fourier Transform (IGFT) is

$$\mathbf{x} = \mathbf{U}\mathbf{s} \tag{3}$$

If a signal is non-zero only within a set of indices $\mathbf{F} \subseteq \{1, \ldots, N\}$ in the frequency domain, it is called band-limited and satisfies

$$\mathbf{s}_i = 0, \forall i \notin \mathbf{F} \tag{4}$$

In this case, the signal can be represented as

$$\mathbf{x} = \mathbf{U}_F\mathbf{s}_F \tag{5}$$

where $\mathbf{U}_F \in \mathbf{R}^{N\times F}$ is the submatrix of eigenvectors corresponding to $\mathbf{F}$, and $\mathbf{s}_F \in \mathbf{R}^{F\times 1}$ are the corresponding non-zero frequency domain coefficients.

Based on the requirements of data acquisition in practical applications, a sampling matrix $\mathbf{D}_\mathbf{S} = \text{diag}(d_1, \ldots, d_N)$ is defined, where $d_i = 1$ indicates that node i is sampled, and 0 otherwise. The number of non-zero entries in $d_i = 1$ corresponds to the number of sampling points $|\mathbf{S}|$ [38], [39].The sampled signal is given by

$$\mathbf{y} = \mathbf{D}_\mathbf{S}\mathbf{x} \tag{6}$$

This work only considers the case where $\mathbf{A}$ is decomposable.

## III. The Proposed Algorithms

### *A. GSP LLAF Algorithm*

First, the LLAF function is analyzed. By defining $e[i] = x[i] - y[i]$ and introducing the design parameter $p$, the LLAF cost function is shown in

$$J_{\text{LLAF}}(e[i]) = \frac{1}{p}\ln(\cosh(pe[i])) \tag{7}$$

The variation of LLAF and its derivative function with error for different values of the parameters are shown in Fig. 1.

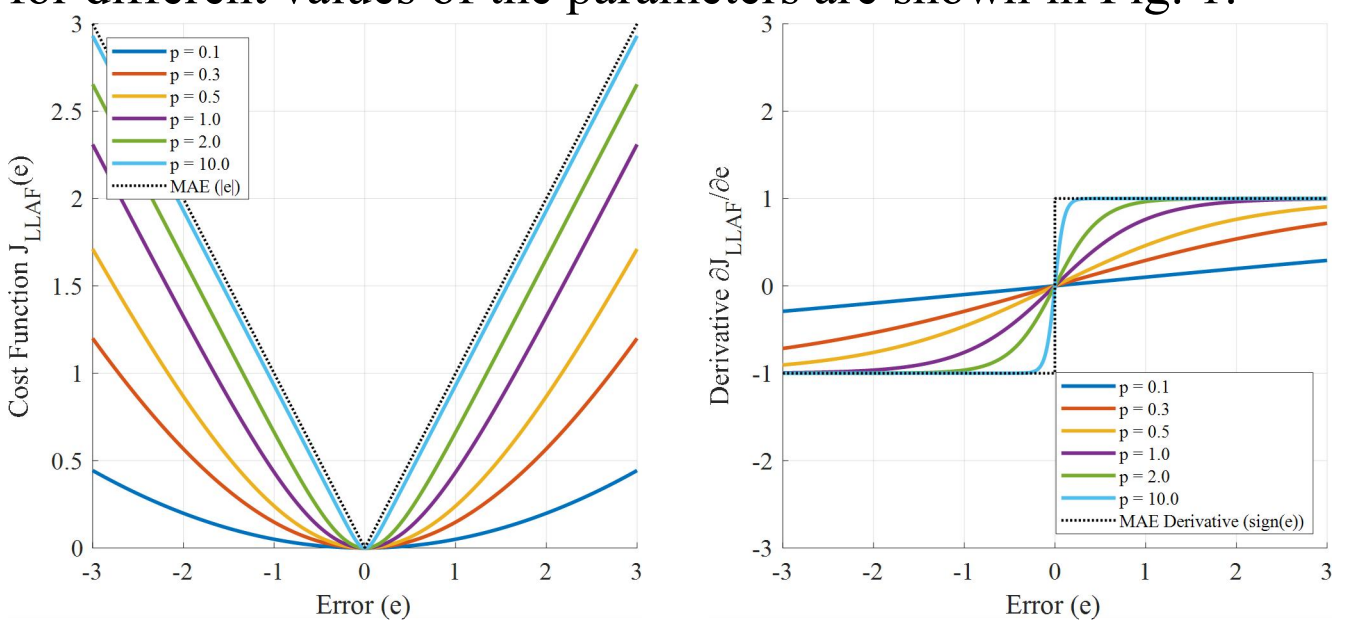

(a) (b)

Fig. 1. Curves of the LLAF function and its derivative versus error.

Fig. 1 shows that, when the $p$ value is small, the LLAF exhibits a scaled mean squared error (MSE) – like shape, whereas when the $p$ value is large it assumes a linear mean absolute error (MAE)–like shape. Moreover, the derivative of the LLAF is bounded, rendering it robust to impulsive and heavy-tailed noise. It also preserves the smoothness property of $L_2$, which facilitates the use of gradient-based optimization methods.

If the error $e$ is finite, then when $p \to 0$, by Taylor expansion we obtain

$$J_{\text{LLAF}} = \frac{1}{p}\left(\frac{p^2e^2}{2} - \frac{p^4e^4}{12} + \cdots\right) = \frac{pe^2}{2} - \frac{p^3e^4}{12} + O\left(p^5\right) \quad (8)$$

$$\frac{\partial J_{\text{LLAF}}}{\partial e} = \tanh(pe) = pe - \frac{p^3e^3}{3} + O\left(p^5\right) \quad (9)$$

Therefore, when $p \to 0$, the values of $J_{\text{LLAF}}$ and $\partial J_{\text{LLAF}}/\partial e$ become negligibly small and thus are ineffective.

When $p \to \infty$, by Taylor expansion we obtain

$$J_{\text{LLAF}} = \frac{1}{p}\left(|pe| - \ln 2 + \cdots\right) = |e| - \frac{\ln 2}{p} + O\left(\frac{e^{-2p|e|}}{p}\right) \quad (10)$$

$$\begin{aligned}\frac{\partial J_{\text{LLAF}}}{\partial e} = \tanh\left(pe\right) &= \begin{cases} 1 - 2e^{-2pe} + O\left(e^{-4p|e|}\right), & e > 0 \\ 2e^{2pe} - 1 + O\left(e^{-4p|e|}\right), & e < 0 \end{cases} \\ &= \operatorname{sign}\left(e\right) + O\left(e^{-4p|e|}\right)\end{aligned} \quad (11)$$

Therefore, when $p \to \infty$, we have $J_{\text{LLAF}} \to |e|$ and $\partial J_{\text{LLAF}}/\partial e \to \operatorname{sign}\left(e\right)$. The LLAF reduces to the Mean Absolute Error (MAE), and the GSP LLAF algorithm reduces to the GSP Sign algorithm [32].

Consider the graph signal to be estimated with N nodes as $\mathbf{x_o} \in \mathbf{R}^{N\times 1}$, and the noise as $\mathbf{w} \in \mathbf{R}^{N\times 1}$. The observation model at the i-th time instant can be defined as

$$\mathbf{x_w}\left[i\right] = \mathbf{x_o}\left[i\right] + \mathbf{w}\left[i\right] \quad (12)$$

Based on (7) and (12), the cost function for the GSP LLAF algorithm is given by

$$\begin{aligned}\mathbf{J}_{\text{GSP LLAF}}\left(\hat{\mathbf{s}}_F\left[i\right]\right) &= \frac{1}{p}\ln\left(\cosh\left(p\mathbf{e}\left[i\right]\right)\right) \\ &= \frac{1}{p}\ln\left(\cosh\left(p\mathbf{D_S}\left(\mathbf{x_w}\left[i\right] - \hat{\mathbf{x}}_\mathbf{o}\left[i\right]\right)\right)\right) \\ &= \frac{1}{p}\ln\left(\cosh\left(p\mathbf{D_S}\left(\mathbf{x_w}\left[i\right] - \mathbf{U}_F\hat{\mathbf{s}}_F\left[i\right]\right)\right)\right)\end{aligned} \quad (13)$$

where $\hat{\mathbf{x}}_\mathbf{o}\left[i\right] = \mathbf{U}_F\hat{\mathbf{s}}_F\left[i\right]$ is the estimated graph signal, and $\mathbf{e}\left[i\right] = \mathbf{D_S}\left(\mathbf{x_w}\left[i\right] - \hat{\mathbf{x}}_\mathbf{o}\left[i\right]\right)$ is the definition of the error.

Taking the gradient of $\mathbf{J}_{\text{GSP LLAF}}\left(\hat{\mathbf{s}}_F\left[i\right]\right)$, we obtain

$$\begin{aligned}\nabla\mathbf{J}_{\text{GSP LLAF}}\left(\hat{\mathbf{s}}_F\left[i\right]\right) &= \frac{\partial \mathbf{J}_{\text{GSP LLAF}}}{\partial \hat{\mathbf{s}}_F\left[i\right]} = \left(\frac{\partial \mathbf{e}\left[i\right]}{\partial \hat{\mathbf{s}}_F\left[i\right]}\right)^T \frac{\partial \mathbf{J}_{\text{GSP LLAF}}}{\partial \mathbf{e}\left[i\right]} \\ &= \left(-\mathbf{D_S}\mathbf{U}_F\right)^T \tanh\left(p\mathbf{e}\left[i\right]\right) \\ &= -\mathbf{U}_F^T \tanh\left(p\mathbf{e}\left[i\right]\right)\end{aligned} \quad (14)$$

According to stochastic gradient descent, the update function for the GSP LLAF algorithm is derived as

$$\begin{aligned}\hat{\mathbf{x}}_\mathbf{o}\left[i+1\right] &= \mathbf{U}_F\left(\hat{\mathbf{s}}_F\left[i\right] - \mu\frac{\partial \mathbf{J}_{\text{GSP LLAF}}}{\partial \hat{\mathbf{s}}_F\left[i\right]}\right) \\ &= \hat{\mathbf{x}}_\mathbf{o}\left[i\right] + \mu\mathbf{U}_F\mathbf{U}_F^T\tanh\left(p\mathbf{e}\left[i\right]\right)\end{aligned} \quad (15)$$

### *B. GSP MVSS-LLAF Algorithm*

To strike a balance between convergence and estimation accuracy, we propose in this section a variable-step-size variant of GSP LLAF, termed the GSP MVSS-LLAF algorithm [40]. The step-size update function for the GSP MVSS-LLAF algorithm is given by

$$\begin{aligned}\mu[i+1] &= \alpha\mu[i] \\ &+ (1-\alpha)\min\left(\frac{\max\left(e_{ng}(i) - \left(\zeta|\mathbf{S}| + 2(1-\zeta)\right)\beta[i], 0\right)}{\left\|\mathbf{U}_F\mathbf{U}_F^T\tanh\left(p\mathbf{e}[i]\right)\right\|_2 + \varepsilon}, \mu[i]\right)\end{aligned} \quad (16)$$

where $\varepsilon > 0$ is a small regularization constant used to avoid division by zero, $\alpha$ is the smoothing exponent, conventionally maintaining $\alpha = 0.9$, and $e_{ng}$ is the mixed error term, expressed as

$$e_{ng}\left[i\right] = \zeta\left\|\mathbf{e}\left[i\right]\right\|_1 + 2\left(1-\zeta\right)\left\|\mathbf{e}\left[i\right]\right\|_2 \quad (17)$$

$\beta\left(i\right)$ is the estimate of the noise absolute value, calculated via the update function

$$\beta\left[i+1\right] = \lambda\beta\left[i\right] + \left(1-\lambda\right)\frac{1}{|\mathbf{S}|}\sum_{k\in\mathbf{S}}\left|\mathbf{e}_k\left[i\right]\right| \quad (18)$$

and $\beta\left[0\right] = 0$. The parameter $\alpha$ is the step-size update smoothing factor, $\zeta$ is the mixing factor, and $\lambda$ is the noise estimate forgetting factor. $\frac{1}{|\mathbf{S}|}\sum_{k\in\mathbf{S}}\left|\mathbf{e}_k\left[i\right]\right|$ denotes taking the absolute error value of the $|\mathbf{S}|$ sampled points from the error vector $\mathbf{e}\left[i\right]$ at time instant *i*.

---

**Algorithm:** GSP MVSS-LLAF

---

**Initialization:** $\hat{\mathbf{x}}_\mathbf{o}\left[0\right] = \mathbf{0}, \beta\left[0\right] = 0$

**Algorithmic parameter:** $p, \alpha, \zeta, \lambda$

choose $\mu\left[0\right] = \mu_0$

**Algorithm implementation:**

**For each iteration *i***

$\mathbf{e}\left[i\right] = \mathbf{D_S}\left(\mathbf{x_w}\left[i\right] - \hat{\mathbf{x}}_\mathbf{o}\left[i\right]\right)$

$e_{ng}\left[i\right] = \zeta\left\|\mathbf{e}\left[i\right]\right\|_1 + 2\left(1-\zeta\right)\left\|\mathbf{e}\left[i\right]\right\|_2$

$\beta\left[i+1\right] = \lambda\beta\left[i\right] + \left(1-\lambda\right)\frac{1}{|\mathbf{S}|}\sum_{k\in\mathbf{S}}\left|\mathbf{e}_k\left[i\right]\right|$

$\mu\left[i+1\right] = \alpha\mu\left[i\right] + (1-\alpha)\min\left(\frac{\max\left(e_{ng}\left[i\right] - \left(\zeta|\mathbf{S}| + 2(1-\zeta)\right)\beta[i], 0\right)}{\left\|\mathbf{U}_F\mathbf{U}_F^T\tanh\left(p\mathbf{e}[i]\right)\right\|_2 + \varepsilon}, \mu\left[i\right]\right)$

$\hat{\mathbf{x}}_\mathbf{o}\left[i+1\right] = \hat{\mathbf{x}}_\mathbf{o}\left[i\right] + \mu\mathbf{U}_F\mathbf{U}_F^T\tanh\left(p\mathbf{e}\left[i\right]\right)$

**End**

---

## IV. Algorithms Performance Analysis

### A. GSP LLAF Algorithm Performance Analysis

This section presents a theoretical performance analysis of the GSP LLAF algorithm, covering its mean-square stability, steady-state error analysis, mean absolute deviation analysis, and computational complexity analysis.

Define the estimation error

$$\tilde{\mathbf{x}}_{\mathbf{o}}[i]=\hat{\mathbf{x}}_{\mathbf{o}}[i]-\mathbf{x}_{\mathbf{o}}[i] \tag{19}$$

Its representation in the frequency domain is

$$\tilde{\mathbf{s}}_F[i]=\mathbf{U}_F^T\tilde{\mathbf{x}}_{\mathbf{o}}[i] \tag{20}$$

Furthermore, in conducting the performance analysis, we assume that:

Assumption 1: The graph signal $\mathbf{x}$ is bandwidth limited.

Assumption 2: This sampling matrix corresponds to a constant sampling set, denoted as $\mathbf{D}_\mathrm{S}(i)=\mathbf{D}_\mathrm{S}$.

Assumption 3: The noise is independently and identically distributed (i.i.d.) and symmetric [31].

Assumption 4: $\tilde{\mathbf{s}}_F[i]$ and $\mathbf{w}[i]$ are mutually independent and uncorrelated [41].

When the noise distribution possesses finite second-order moments, we employ the following assumptions for the Mean Square Stability Analysis and Steady-State Error Analysis:

Assumption 5: In the context of MSD analysis, the noise $\mathbf{w}[i]$ has finite variance.

For noise with infinite variance, this paper adopts the mean absolute deviation (MAD) to analyze the corresponding estimation performance, under the following conditions:

Assumption 6: For performance evaluation, the difference between the estimate $\hat{\mathbf{x}}_{\mathbf{o}}[i]$ and the noise-free target $\mathbf{x}_{\mathbf{o}}[i]$ is available.

#### 1) Mean Square Stability Analysis

In this section, we present the mean-square convergence condition for the GSP LLAF algorithm. Under the common assumption of bounded second-order moments, this condition implies first-order convergence. As second-order recursion involves higher-order moments of the stochastic gradient, the step-size upper bound for mean-square convergence is typically more stringent than that for mean convergence. Therefore, this paper primarily uses the mean-square condition as a reference for step-size design.

From (15) and (19), the estimation error update equation is obtained

$$\begin{aligned}\tilde{\mathbf{x}}_{\mathbf{o}}[i+1]&=\hat{\mathbf{x}}_{\mathbf{o}}[i+1]-\mathbf{x}_{\mathbf{o}}[i]\\&=\tilde{\mathbf{x}}_{\mathbf{o}}[i]+\mu\mathbf{U}_F\mathbf{U}_F^T\tanh\left(p\mathbf{D}_\mathrm{S}\left(\mathbf{x}_{\mathbf{o}}[i]+\mathbf{w}[i]-\hat{\mathbf{x}}_{\mathbf{o}}[i]\right)\right)\\&=\tilde{\mathbf{x}}_{\mathbf{o}}[i]+\mu\mathbf{U}_F\mathbf{U}_F^T\tanh\left(p\mathbf{D}_\mathrm{S}\left(\mathbf{w}[i]-\tilde{\mathbf{x}}_{\mathbf{o}}[i]\right)\right)\end{aligned} \tag{21}$$

Multiplying both sides of (21) on the left by $\mathbf{U}_F^T$ yields its representation in the frequency domain

$$\tilde{\mathbf{s}}_F[i+1]=\tilde{\mathbf{s}}_F[i]+\mu\mathbf{U}_F^T\tanh\left(p\mathbf{D}_\mathrm{S}\left(\mathbf{w}[i]-\mathbf{U}_F\tilde{\mathbf{s}}_F[i]\right)\right) \tag{22}$$

Here we define

$$\mathbf{h}[i]=\tanh\left(p\mathbf{D}_\mathrm{S}\left(\mathbf{w}[i]-\mathbf{U}_F\tilde{\mathbf{s}}_F[i]\right)\right) \tag{23}$$

Based on Assumption 4, and given that $\mathbf{w}[i]\gg\tilde{\mathbf{s}}_F[i]$ when $i$ is sufficiently large, we perform a first-order Taylor expansion [42] of $\mathbf{h}[i]$ at the point $\tilde{\mathbf{s}}_F[i]=0$

$$\mathbf{h}[i]\approx\tanh\left(p\mathbf{D}_\mathrm{S}\mathbf{w}[i]\right)-p\mathbf{D}_\mathrm{S}\mathbf{U}_F\tilde{\mathbf{s}}_F[i]\odot\mathrm{sech}^2\left(p\mathbf{D}_\mathrm{S}\mathbf{w}[i]\right) \tag{24}$$

Let $\mathbf{R}_p=\mathrm{diag}\left(\mathrm{E}\left[\mathrm{sech}^2\left(p\mathbf{D}_\mathrm{S}\mathbf{w}[i]\right)\right]\right)$ be the expected gain matrix. Due to Assumption 2 and Assumption 4, we have $\mathbf{R}_p=r_p\mathbf{D}_\mathrm{S}+(\mathbf{I}-\mathbf{D}_\mathrm{S})$, where $r_p=\mathrm{E}\left[\mathrm{sech}^2(pw)\right]$, and $w$ is a scalar random variable denoting the noise component $\mathbf{w}[i]$ at a single node.

Substituting (24) into (22) yields

$$\begin{aligned}\tilde{\mathbf{s}}_F[i+1]&\approx\tilde{\mathbf{s}}_F[i]\\&\quad+\mu\mathbf{U}_F^T\mathbf{D}_\mathrm{S}\left[\tanh\left(p\mathbf{D}_\mathrm{S}\mathbf{w}[i]-p\mathbf{D}_\mathrm{S}\mathbf{U}_F\tilde{\mathbf{s}}_F[i]\cdot r_p\right)\right]\\&\approx\left(\mathbf{I}-\mu pr_p\mathbf{U}_F^T\mathbf{D}_\mathrm{S}\mathbf{U}_F\right)\tilde{\mathbf{s}}_F[i]\\&\quad+\mu\mathbf{U}_F^T\mathbf{D}_\mathrm{S}\tanh\left(p\mathbf{D}_\mathrm{S}\mathbf{w}[i]\right)\end{aligned} \tag{25}$$

Let $\mathbf{G}=\mathbf{U}_F^T\mathbf{D}_\mathrm{S}\mathbf{U}_F$, $\mathbf{v}[i]=\tanh\left(p\mathbf{D}_\mathrm{S}\mathbf{w}[i]\right)$, then (25) can be written as

$$\begin{aligned}\tilde{\mathbf{s}}_F[i+1]&\approx\tilde{\mathbf{s}}_F[i]\\&\quad+\mu\mathbf{U}_F^T\mathbf{D}_\mathrm{S}\left[\tanh\left(p\mathbf{D}_\mathrm{S}\mathbf{w}[i]-p\mathbf{D}_\mathrm{S}\mathbf{U}_F\tilde{\mathbf{s}}_F[i]\cdot r_p\right)\right]\\&\approx\left(\mathbf{I}-\mu pr_p\mathbf{G}\right)\tilde{\mathbf{s}}_F[i]+\mu\mathbf{U}_F^T\mathbf{D}_\mathrm{S}\mathbf{v}[i]\end{aligned} \tag{26}$$

The mean-square deviation is defined as

$$\mathrm{MSD}[i]=\mathrm{E}\left\|\tilde{\mathbf{s}}_F[i]\right\|_2^2 \tag{27}$$

Using the MSD definition in (27), we take the expectation of the squared Euclidean norm of the error recursion in (26). Under Assumption 4, the cross term has zero expectation, yielding

$$\begin{aligned}\mathrm{MSD}[i+1]&=\mathrm{E}\left\|\tilde{\mathbf{s}}_F[i+1]\right\|_2^2\\&\approx\mathrm{E}\left\|\left(\mathbf{I}-\mu pr_p\mathbf{G}\right)\tilde{\mathbf{s}}_F[i]\right\|_2^2\\&\quad+\mu^2\mathrm{E}\left\|\mathbf{U}_F^T\mathbf{D}_\mathrm{S}\mathbf{v}[i]\right\|_2^2\end{aligned} \tag{28}$$

Since $\mathbf{v}[i]=\tanh\left(p\mathbf{D}_\mathrm{S}\mathbf{w}[i]\right)$, and $\tanh(\cdot)$ is an odd function, it follows from Assumption 3 that $\mathrm{E}\left[\mathbf{v}[i]\right]=\mathbf{0}$, and

$$\mathrm{E}\left[\mathbf{v}[i]\mathbf{v}[i]^T\right]=\sigma_v^2\mathbf{D}_\mathrm{S} \tag{29}$$

where $\sigma_v^2=\mathrm{E}\left[\tanh^2(pw)\right]$.

Based on (29), (28) can be expanded as

$$\begin{aligned}\mathrm{E}\left\|\tilde{\mathbf{s}}_F[i+1]\right\|^2&=\mathrm{E}\left[\tilde{\mathbf{s}}_F[i]^T\left(\mathbf{I}-\mu pr_p\mathbf{G}\right)^T\left(\mathbf{I}-\mu pr_p\mathbf{G}\right)\tilde{\mathbf{s}}_F[i]\right]\\&\quad+\mu^2\mathrm{Tr}\left[\mathbf{U}_F^T\mathbf{D}_\mathrm{S}\mathrm{E}\left[\mathbf{v}[i]\mathbf{v}[i]^T\right]\mathbf{D}_\mathrm{S}\mathbf{U}_F\right]\\&=\mathrm{E}\left[\tilde{\mathbf{s}}_F[i]^T\left(\mathbf{I}-\mu pr_p\mathbf{G}\right)^T\left(\mathbf{I}-\mu pr_p\mathbf{G}\right)\tilde{\mathbf{s}}_F[i]\right]\\&\quad+\mu^2\sigma_v^2\mathrm{Tr}(\mathbf{G})\end{aligned} \tag{30}$$

The error covariance matrix is defined as

$$\mathbf{P}[i]=\mathrm{E}\left[\tilde{\mathbf{s}}_F[i]\tilde{\mathbf{s}}_F[i]^T\right] \tag{31}$$

To derive the covariance recursion, let $\mathbf{A} = \mathbf{I} - \mu p r_p \mathbf{G}$ and rewrite (26) as $\tilde{s}_F[i+1] = \mathbf{A}\tilde{s}_F[i] + \mu \mathbf{U}_F^T \mathbf{D}_S \mathbf{v}[i]$. Taking the outer product of both sides and then the expectation, the cross terms vanish under Assumptions 3 and 4. Therefore, $\mathbf{P}[i+1] = \mathbf{A}\mathbf{P}[i]\mathbf{A}^T + \mu^2 \mathbf{U}_F^T \mathbf{D}_S \mathrm{E}\left[\mathbf{v}[i]\mathbf{v}^T[i]\right]\mathbf{D}_S \mathbf{U}_F$. Using $\mathrm{E}\left[\mathbf{v}[i]\mathbf{v}^T[i]\right] = \sigma_v^2 \mathbf{D_S}$, $\mathbf{D}_S^2 = \mathbf{D}_S$, and $\mathbf{G} = \mathbf{U}_F^T \mathbf{D}_S \mathbf{U}_F$, the covariance recursion becomes

$$\mathbf{P}[i+1] = \left(\mathbf{I} - \mu p r_p \mathbf{G}\right)\mathbf{P}[i]\left(\mathbf{I} - \mu p r_p \mathbf{G}\right)^T + \mu^2 \sigma_v^2 \mathbf{G} \quad (32).$$

When the algorithm reaches steady state, we have $\lim_{i\to\infty} \tilde{\mathbf{s}}_F[i] = \lim_{i\to\infty} \tilde{\mathbf{s}}_F[i+1]$ and $\mathbf{P}[i+1] = \mathbf{P}[i] = \mathbf{P}_\infty$, satisfying the Lyapunov equation [14]

$$\mathbf{P}_\infty = \left(\mathbf{I} - \mu p r_p \mathbf{G}\right)\mathbf{P}_\infty\left(\mathbf{I} - \mu p r_p \mathbf{G}\right)^T + \mu^2 \sigma_v^2 \mathbf{G} \quad (33)$$

A necessary and sufficient condition for algorithm convergence is that the spectral radius of matrix $\left(\mathbf{I} - \mu p r_p \mathbf{G}\right)$ is less than 1, denoted as

$$\left|1 - \mu p r_p \lambda_k(\mathbf{G})\right| < 1, \forall k \quad (34)$$

where $\lambda_k(\mathbf{G})$ are the eigenvalues of $\mathbf{G}$. This yields the step-size constraint condition

$$0 < \mu < \frac{2}{p r_p \lambda_{\max}(\mathbf{G})} \quad (35)$$

*2) Steady-State Error Analysis*

According to (27) and (31), we have

$$\mathrm{MSD}[i] = \mathrm{Tr}\left(\mathbf{P}[i]\right) \quad (36)$$

The steady-state MSD is then

$$\mathrm{MSD}_\infty = \mathrm{Tr}\left(\mathbf{P}_\infty\right) \quad (37)$$

According to (33) and (25), when $\mathbf{G} = \mathbf{U}_F^T \mathbf{D_S} \mathbf{U}_F$ is diagonalizable, (37) has a closed-form solution. Let $\mathbf{G} = \mathbf{V}\boldsymbol{\Pi}\mathbf{V}^T$, where $\mathbf{V}$ is an orthogonal matrix and $\lambda_k > 0$ is the *k*-th eigenvalue of $\mathbf{G}$. Define the transformed steady-state covariance matrix as

$$\overline{\mathbf{P}}_\infty = \mathbf{V}^T \mathbf{P}_\infty \mathbf{V} \quad (38)$$

Premultiplying and postmultiplying (32) by $\mathbf{V}^T$ and $\mathbf{V}$, respectively, yields

$$\overline{\mathbf{P}}_\infty = \left(\mathbf{I} - \mu p r_p \boldsymbol{\Pi}\right)\overline{\mathbf{P}}_\infty\left(\mathbf{I} - \mu p r_p \boldsymbol{\Pi}\right)^T + \mu^2 \sigma_v^2 \boldsymbol{\Pi} \quad (39)$$

Because $\boldsymbol{\Pi}$ is diagonal, its *k*-th diagonal element satisfies

$$\left[\overline{\mathbf{P}}_\infty\right]_{kk} = \left(\mathbf{I} - \mu p r_p \lambda_k\right)\left[\overline{\mathbf{P}}_\infty\right]_{kk} + \mu^2 \sigma_v^2 \lambda_k \quad (40)$$

Solving for $\left[\overline{\mathbf{P}}_\infty\right]_{kk}$ and using the invariance of the trace under an orthogonal transformation, we obtain

$$\mathrm{MSD}_\infty = \mathrm{Tr}\left(\mathbf{P}_\infty\right) = \mu^2 \sigma_v^2 \sum_{k=1}^{F} \frac{\lambda_k}{1 - \left(1 - \mu p r_p \lambda_k\right)^2} = \frac{\mu \sigma_v^2}{p r_p} \sum_{k=1}^{F} \frac{1}{2 - \mu p r_p \lambda_k} \quad (41)$$

When the step size is sufficiently small, with $\mu p r_p \lambda_k \ll 1$ holding, (41) can be further approximated as

$$\mathrm{MSD}_\infty \approx \frac{\mu F \sigma_v^2}{2 p r_p} \quad (42)$$

*3) Mean Absolute Deviation Analysis*

The MSD analyses in the preceding subsections are restricted to noise distributions with a finite second-order moment. For noise with infinite variance, such as Cauchy noise and symmetric α-stable noise with $\alpha < 2$, the corresponding estimation performance is analyzed using the mean absolute deviation (MAD).

The node-averaged MAD is defined as

$$\mathrm{MAD}[i] = \frac{1}{N}\mathrm{E}\left[\left\|\hat{\mathbf{x}}_\mathbf{o}[i] - \mathbf{x}_\mathbf{o}\right\|_1\right] = \frac{1}{N}\mathrm{E}\left[\left\|\tilde{\mathbf{x}}_\mathbf{o}[i]\right\|_1\right] \quad (43)$$

Using the linearized frequency-domain error recursion in (26), we have

$$\tilde{\mathbf{s}}_F[i+1] \approx \mathbf{A}\tilde{\mathbf{s}}_F[i] + \mu \mathbf{U}_F^T \mathbf{D}_S \mathbf{v}[i] \quad (44)$$

where $\mathbf{A} = \mathbf{I} - \mu p r_p \mathbf{G}$, $\mathbf{v}[i] = \tanh\left(p \mathbf{D}_S \mathbf{w}[i]\right)$, and $\mathbf{G} = \mathbf{U}_F^T \mathbf{D}_S \mathbf{U}_F$. Taking the Euclidean norm of both sides of (44), applying the triangle inequality, and then taking the expectation yield

$$\mathrm{E}\left[\left\|\tilde{\mathbf{s}}_F[i+1]\right\|\right]_2 \le \rho \mathrm{E}\left[\left\|\tilde{\mathbf{s}}_F[i]\right\|\right]_2 + \mu c_v \quad (45)$$

where $\rho = \left\|\mathbf{I} - \mu p r_p \mathbf{G}\right\|_2$, $c_v = \mathrm{E}\left[\left\|\mathbf{U}_F^T \mathbf{D}_S \mathbf{v}[i]\right\|_2\right]$.

Since $\tanh(a) \le 1$ for any $a$, the transformed-noise term is bounded even when $\mathbf{w}[i]$ has infinite variance. In particular,

$$c_v \le \left\|\mathbf{U}_F^T\right\|_2 \mathrm{E}\left[\left\|\mathbf{D}_S \mathbf{v}[i]\right\|_2\right] \le \sqrt{|\mathbf{S}|} \quad (46)$$

Recursively applying (45) gives

$$\mathrm{E}\left[\left\|\tilde{s}_F[i]\right\|_2\right] \le \rho^i \mathrm{E}\left[\left\|\tilde{s}_F[0]\right\|_2\right] + \mu c_v \frac{1 - \rho^i}{1 - \rho} \quad (47)$$

For a qualified sampling set, $\mathbf{G}$ is positive definite, and $\rho = \max_k \left|1 - \mu p r_p \lambda_k(\mathbf{G})\right|$. Therefore, the step-size condition in (35) ensures that $\rho < 1$. Furthermore, because $\tilde{\mathbf{x}}_\mathbf{o}[i] = \mathbf{U}_F \tilde{\mathbf{s}}_F[i]$ and $\mathbf{U}_F$ has orthonormal columns, $\frac{1}{N}\left\|\tilde{x}_o[i]\right\|_1 \le \frac{1}{\sqrt{N}}\left\|\tilde{\mathbf{s}}_F[i]\right\|_2$. Combining this relation with (47), we obtain

$$\mathrm{MAD}[i] \le \frac{\rho^i}{\sqrt{N}} \mathrm{E}\left[\left\|\tilde{s}_F[0]\right\|_2\right] + \frac{\mu c_v}{\sqrt{N}} \frac{1 - \rho^i}{1 - \rho} \quad (48)$$

Consequently, the steady-state MAD satisfies the following upper bound:

$$\limsup_{i\to\infty} \mathrm{MAD}[i] \le \frac{\mu c_v}{\sqrt{N}(1 - \rho)} \le \frac{\mu \sqrt{|\mathbf{S}|}}{\sqrt{N}(1 - \rho)} \quad (49)$$

The above conclusion does not require $\mathrm{E}\left[w^2\right]$ or $\mathrm{E}\left[|w|\right]$ to be finite. This is because both $\tanh(pw)$ and $\mathrm{sech}^2(pw)$ are bounded functions, which ensures that $c_v$ and $r_p$ remain finite even when the noise has infinite variance.

Therefore, the GSP LLAF algorithm still possesses a finite upper bound on the first-order steady-state error in this case.

*4) Computational Complexity Analysis*

In this section we evaluate and contrast the per-iteration computational burden of the proposed GSP LLAF and several benchmark algorithms. The accounting covers elementary arithmetic and matrix operations relevant to the update stage, namely additions/subtractions, multiplications/divisions, power multiplications, exponentials, and direct matrix inversion (DMI). All complexity measures are restricted to the update block of each algorithm's graph-signal adaptive estimation routine; for instance, the update block for the GSP-LLAF algorithm corresponds to entry " $\mathbf{U}_F\mathbf{U}_F^T \tanh\left(p\mathbf{e}[i]\right)$ " in our notation, and analogous references apply to the other methods.

TABLE I
ANALYSIS OF THE ACTUAL COMPUTATIONAL COMPLEXITY OF THE RELATED OPERATIONS

| Operation | Multiplication | Addition |
|---|---|---|
| $\mathbf{U}_F^T\mathbf{e}(i)$ | $F\lvert\mathbf{S}\rvert$ | $F(\lvert\mathbf{S}\rvert-1)$ |
| $\mathbf{D}_\mathbf{S}\mathbf{U}_F$ | $F\lvert\mathbf{S}\rvert$ | 0 |
| $\mathbf{U}_F^T\mathbf{D}_\mathbf{S}\mathbf{U}_F$ | $F^2\lvert\mathbf{S}\rvert$ | $F^2(\lvert\mathbf{S}\rvert-1)$ |
| $\mathbf{e}^T(i)\ \mathbf{e}(i)$ | $\lvert\mathbf{S}\rvert$ | $\lvert\mathbf{S}\rvert-1$ |
| $\left(\mathbf{U}_F^T\mathbf{D}_\mathbf{S}\mathbf{U}_F\right)^{-1}$ | $\frac{1}{3}F^3$ | 0 |

When a matrix of shape $T_1\times T_2$ is multiplied with a matrix of shape $T_2\times T_3$, the computations in both multiplication and addition are $T_1\times T_2\times T_3$ and $T_1\times T_3\times(T_2-1)$, respectively. Some actual computational complexity of the related operations is listed in Table I.

The computational complexity of the $\tanh(\cdot)$ term in the GSP LLAF algorithm in Table II is denoted as $\mathrm{O}(2\lvert\mathbf{S}\rvert)$. From Table II, the following conclusions can be drawn: the computational complexity of the proposed GSP LLAF algorithm is significantly lower than that of algorithms such as GSP MCC and GSP GMCC. Furthermore, compared to the GSP Sign algorithm, its computational complexity has only increased by $\mathrm{O}(2\lvert\mathbf{S}\rvert)$. This increment arises from the multiplication involving the parameter $p$ and the $\tanh(\cdot)$ term in the GSP LLAF algorithm. Therefore, we consider the lower computational complexity to be an advantage of the GSP LLAF algorithm.

*B. GSP MVSS-LLAF Algorithm Performance Analysis*

In this section, the stability of the step-size update term is analyzed. Define

$$q[i]=\frac{\max\left(e_{ng}[i]-C_S\beta[i],0\right)}{\left\|\mathbf{U}_F\mathbf{U}_F^T\tanh\left(p\mathbf{e}[i]\right)\right\|_2+\varepsilon},\quad C_S=\zeta\lvert\mathbf{S}\rvert+2(1-\zeta)\tag{50}$$

Equation (16) can then be written as

$$\mu[i+1]=\alpha\mu[i]+(1-\alpha)\min\left(q[i],\mu[i]\right)\tag{51}$$

Since $q[i]\ge 0$, $\mu[0]\ge 0$, and $0<\alpha<1$, it follows that

$$\alpha\mu[i]\le\mu[i+1]\le\mu[i]\tag{52}$$

Therefore, $\{\mu[i]\}$ is nonnegative and monotonically non-increasing and hence converges to a finite nonnegative value. At steady state, $\beta[i]$ follows the recent sample-based absolute-residual level, and the truncation condition $e_{ng}[i]\le C_S\beta[i]$ is repeatedly activated. At each such iteration $i_k$, $q[i_k]=0$ and $\mu[i_k+1]=\alpha\mu[i_k]$. Consequently, after $K$ activation events,

$$0\le\mu[i_K+1]\le\alpha^K\mu[0]\underset{K\to\infty}{\rightarrow}0\tag{53}$$

Thus, when the algorithm reaches steady state, the step size approaches an extremely small value. In practice, the update can be terminated when $\mu[i]$ falls below a prescribed threshold $\mu_{\text{stop}}$. Therefore, a sufficiently small step size can be regarded as the update stopping criterion.

## V. COMPUTER SIMULATION EXPERIMENT

This section will conduct a discussion of algorithm parameters, algorithm performance comparison under different $\lvert\mathbf{S}\rvert$, verification of the theoretical performance of the algorithms, static temperature graph signal estimation application experiments, and dynamic temperature graph signal prediction application experiments.

Sections A and B employ randomly generated graph signal data $\mathbf{G}_1$ and $\mathbf{G}_2$. To construct $\mathbf{G}_1$ and $\mathbf{G}_2$, $N=100$ node positions $\{\mathbf{p}_i\}_{i=1}^{N}$ are independently and uniformly sampled within the unit square $[0,1]^2$. We define the Euclidean distance $d_{nm}=\|\mathbf{p}_n-\mathbf{p}_m\|_2$ between nodes and a connection radius threshold

$$r=\sqrt{1.2\log N/N} \tag{54}$$

TABLE II
COMPUTATIONAL COMPLEXITY ANALYSIS AND COMPARISON OF THE RELATED ALGORITHMS

| Algorithm | Multiplication/Division | Addition | exp | power product | DMI | $\mathrm{Sign}(\cdot)$ | Overall computational complexity |
|---|---|---|---|---|---|---|---|
| GSP LMS | $F(\lvert\mathbf{S}\rvert+N)$ | $F\lvert\mathbf{S}\rvert-F+NF-N$ | 0 | 0 | 0 | 0 | $2F\lvert\mathbf{S}\rvert+2FN-F-N$ |
| GSP NLMS | $(\lvert\mathbf{S}\rvert+N)(F^2+F)$ | $F^2(\lvert\mathbf{S}\rvert+N-1)+F(\lvert\mathbf{S}\rvert-1)-N$ | 0 | 0 | $\frac{1}{3}F^3$ | 0 | $2F^2\lvert\mathbf{S}\rvert+2F\lvert\mathbf{S}\rvert+2F^2N$ $+NF-F^2-F-N+\frac{1}{3}F^3$ |
| GSP MCC | $N^2(F+1)+N+\lvert\mathbf{S}\rvert$ | $N(NF-1)$ | $N$ | $N$ | 0 | 0 | $2N^2F+N^2+\lvert\mathbf{S}\rvert+2N$ |
| GSP GMCC | $N^2(F+1)+2N+\lvert\mathbf{S}\rvert$ | $N(NF-1)$ | $N$ | $2N$ | 0 | 0 | $2N^2F+N^2+4N+\lvert\mathbf{S}\rvert$ |
| GSP Sign | $F(N+\lvert\mathbf{S}\rvert)+\lvert\mathbf{S}\rvert$ | $F(\lvert\mathbf{S}\rvert-1)+N(F-1)$ | 0 | 0 | 0 | $\lvert\mathbf{S}\rvert$ | $2F\lvert\mathbf{S}\rvert+2FN+2\lvert\mathbf{S}\rvert-F-N$ |
| GSP LLAF | $F(N+\lvert\mathbf{S}\rvert)+2\lvert\mathbf{S}\rvert$ | $F(\lvert\mathbf{S}\rvert-1)+N(F-1)$ | $2\lvert\mathbf{S}\rvert$ | 0 | 0 | 0 | $2F\lvert\mathbf{S}\rvert+2FN+4\lvert\mathbf{S}\rvert-F-N$ |

When $d_{nm}<r$, an edge is established between nodes $n$ and $m$, and a Gaussian kernel is employed to assign a weight to this edge

$$a_{nm}=\exp\left(-\frac{d_{nm}^2}{2\sigma^2}\right),\quad \sigma=\frac{r}{2} \tag{55}$$

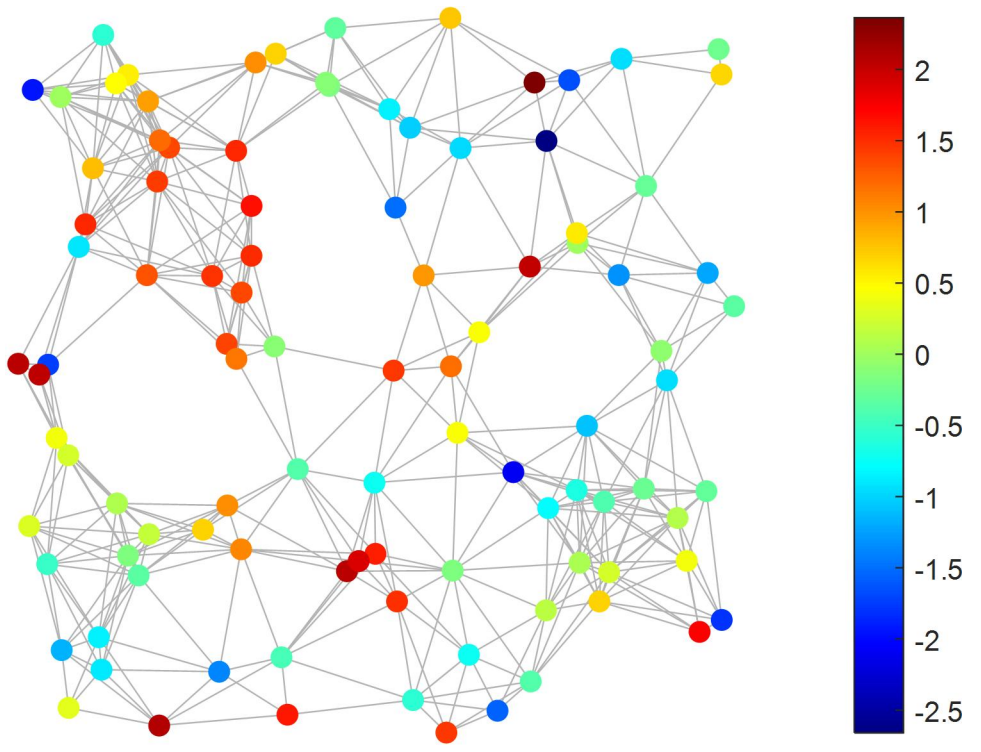

Fig. 2. Randomly generated graph $\mathbf{G}_1$.

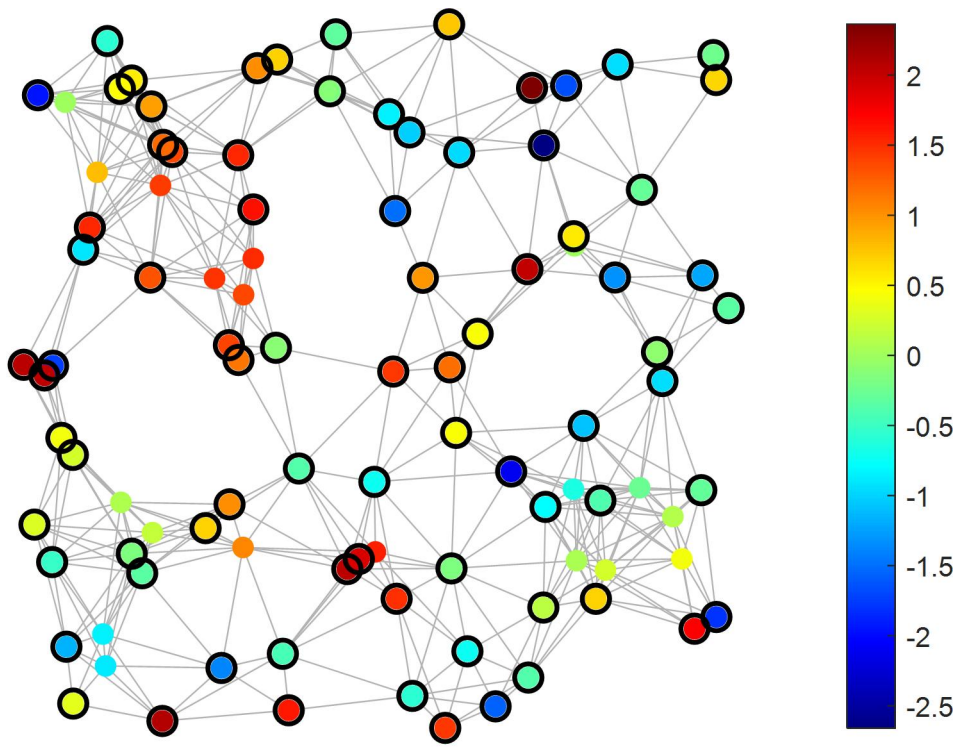

Fig. 3. Schematic diagram of sampling $\mathbf{G}_1$ using the Max-Det sampling strategy.

Otherwise, $a_{nm}=0$, thereby obtaining $\mathbf{A}$. Subsequently, $\mathbf{U}$ is derived based on (1). Taking the first $F$ eigenvectors as the basis for the low-frequency subspace in the frequency domain, the low-frequency basis matrix $\mathbf{U}_F$ is obtained. Construct $\mathbf{s}_F=[s_1,\dots,s_F]^T$, with coefficients independently and identically sampled from a symmetric uniform distribution

$$s_k\sim\mathbf{u}\left(-\frac{\Delta}{2},\frac{\Delta}{2}\right),\quad k=1,\dots,F \tag{56}$$

where $\Delta=\sqrt{12\sigma_s^2}$ and $\sigma_s^2=2$ are predetermined variances.

Finally, the vertex-domain signal $\mathbf{x}$ is obtained from (5). When $F=60$, the generated $\mathbf{G}_1$ is shown in Fig. 2. The sampling strategy uses the Max-Det sampling strategy [14]. Fig. 3 shows the sampling situation for $\mathbf{G}_1$ when $|\mathbf{S}|=80$.

Sections C and D utilize the July temperature data from the Brazilian meteorological temperature dataset [43] to construct the graph $\mathbf{G}_3$, as shown in Fig. 4. The relevant parameters are set as: $N=299, K=8, |\mathbf{S}|=210, F=200$. This dataset contains a total of 299 stations, with each station connected to its eight nearest neighboring stations. Data from 210 stations were collected. The weight $a_{nm}$ between node n and node m in the graph is computed via

$$a_{nm}=\exp\left(-\frac{d_h(n,m)^2}{2\theta^2}\right) \tag{57}$$

where $d_h(n,m)$ is the Haversine distance between the two stations, and $\theta$ is the scale parameter. The graph signal is sampled via the Max-$\lambda_{\min}$ sampling strategy [38], [39], [43]. The Haversine distance is defined as

$$d_h(n,m)=2R_{\mathrm{E}}\arcsin\left(\sqrt{\sin^2\left(\frac{\phi_n-\phi_m}{2}\right)+\cos(\phi_n)\cos(\phi_m)\sin^2\left(\frac{\lambda_n-\lambda_m}{2}\right)}\right) \tag{58}$$

where $R_{\mathrm{E}}$ is the Earth's radius, and $(\phi_n,\lambda_n)$ and $(\phi_m,\lambda_m)$ are the latitude and longitude coordinates, in radians, of nodes $n$ and m, respectively.

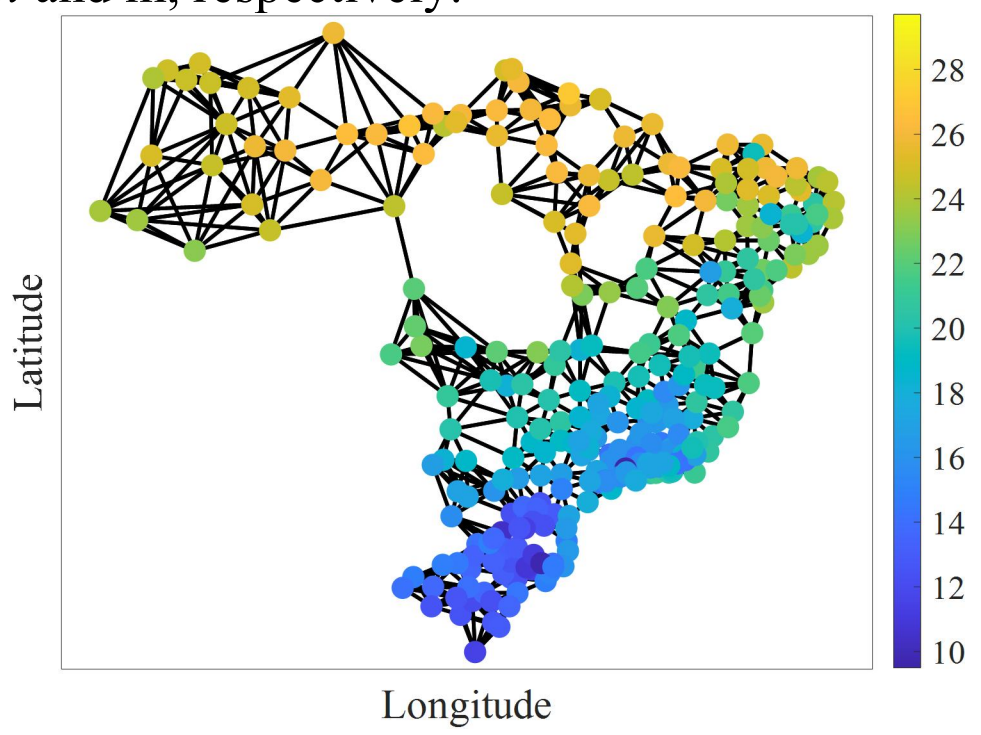

Fig. 4. July temperature data graph $\mathbf{G}_3$ from the Brazilian dataset (Unit: °C).

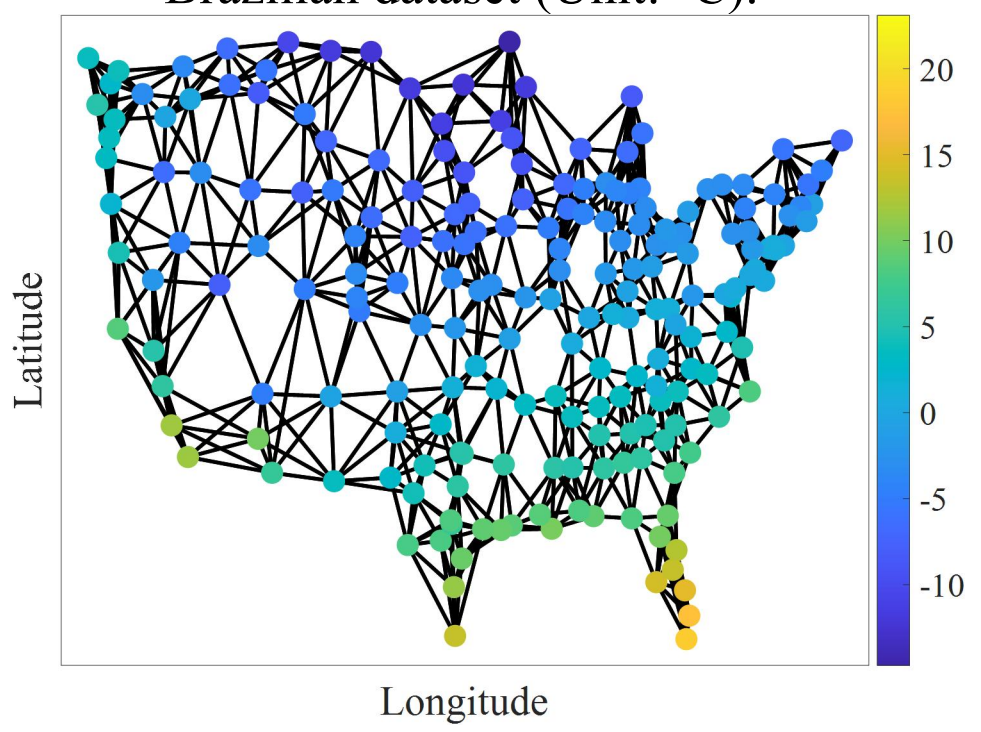

Fig. 5. First-hour temperature data graph $\mathbf{G}_4$ from the U.S. dataset (Unit: °C).

Section E utilizes the U.S. weather temperature dataset [44] to construct the graph $\mathbf{G}_4$, where $\mathbf{G}_4$ is a time-varying streaming graph signal, and the weight $a_{nm}$ is also constructed using (57), with a duration of 95 hours. Fig. 5 depicts the graph signal formed by the temperature data from the first hour in the dataset ( $\mathbf{G}_4$ image at the first hour). The relevant parameters are set as: $K=7,\ N=205,\ |\mathbf{S}|=130,\ F=125$ . This dataset contains a total of 205 stations, with each station connected to its seven nearest neighboring stations. Data from 130 stations were collected. The graph signal is sampled via the sampling strategy [38], [39], [43].

In Experiments 7 and 8, the temperature datasets provide realistic graph structures and signal variations, rather than assuming that meteorological measurement errors exactly follow the adopted distributions. Nevertheless, sensor faults, transmission errors, and interference can produce outliers [45]. Therefore, BG, Laplace, α-stable, and Cauchy noises are introduced as controlled impulsive or heavy-tailed robustness tests. These models are also relevant to communication, state-space inference, and image-restoration applications [42], [46], [47], [48].

In addition, the parameter $\mathrm{MSD_{dB}}$ is used to quantitatively evaluate the performance of the relevant algorithms in terms of convergence/tracking rate and steady state error, which is shown in

$$\mathrm{MSD_{dB}}=10\log_{10}\left\|\hat{\mathbf{x}}_{\mathbf{o}}(i)-\mathbf{x}_{\mathbf{o}}\right\|_2^2 \tag{59}$$

The experiments in this section were conducted on a computer with a 13th Gen Intel(R) Core(TM) i7-13620H CPU and 16GB of RAM, using MATLAB R2024b. For each independent trial in the experiments, the random number generation process was controlled by setting a random number seed via rng(seed). Unless otherwise stated, all simulation results presented were obtained as averages over $M=100$ independent repeated experiments.

The experiments are conducted under α-stable noise, impulse noise simulated using the Bernoulli-Gaussian (BG) model, Laplace noise, and Cauchy noise environments. The noise parameters are set as follows: in the α-stable noise, $\alpha$ denotes the stability index, and $\gamma$ denotes the scale parameter [19]. In the BG impulsive noise, $Pr$ represents the impulse occurrence probability, $\delta_g^2$ denotes the variance of the Gaussian process, and $\delta_b^2$ denotes the variance of the impulsive Gaussian component [46]. $b_L$ indicates the scale parameter of the Laplace noise [47]. $b_C$ indicates the scale parameter of the Cauchy noise [48].

*A. Parameters Discussion*

*1) GSP LLAF Algorithm Parameter Discussion*

This section investigates the impact of different parameters on the performance of the GSP LLAF algorithm through Experiment 1, using the data from $\mathbf{G}_1$ . The remaining parameter settings are listed in Table III. The experiment is conducted under α-stable noise conditions. The simulation results of Experiment 1 are shown in Fig. 6. Finally, the empirical rule for initializing parameters $p$ and $\mu$ in practical applications is presented.

From Fig. 6(a), the following conclusions can be drawn: as the parameter $p$ decreases, the convergence speed of the GSP LLAF algorithm gradually slows down, while the steady-state error gradually decreases. When the parameter $p$ is relatively large (specifically $p>5$ for Experiment 1-1), the steady-state error of the GSP LLAF algorithm slightly increases as the parameter $p$ grows, but the overall performance remains approximately similar. The 3D surface in Fig. 6(b) visually illustrates the parameter coupling and unstable regions. Based on the experimental results, to balance the performance metrics of convergence speed and steady-state error, this section determines the parameter scheme for the GSP LLAF algorithm as $p=0.3, \mu=0.3$ . This parameter scheme will be used in the experiment 3 in the subsequent subsections. Meanwhile, in the multi-algorithm performance comparison experiments, all algorithms adopt step sizes that achieve comparable convergence speeds.

TABLE III
PARAMETER SETTINGS RELATED TO EXPERIMENT 1

| Experiment number | The related parameters | noise environment |
|---|---|---|
| Experiment 1-1 | $\mu=0.3$<br>$p\in\{0.0005,0.005,0.05,0.1,0.3,0.5,0.7,1,5,10\}$ | $\alpha=1.4$<br>$\gamma=1$ |
| Experiment 1-2 | $\mu\in\{0.001,0.01,0.1,0.3,0.5,0.8,1,2,5\}$<br>$p\in\{0.0005,0.005,0.05,0.1,0.3,0.5,0.7,1,5,10\}$ | $\alpha=1.4$<br>$\gamma=1$ |

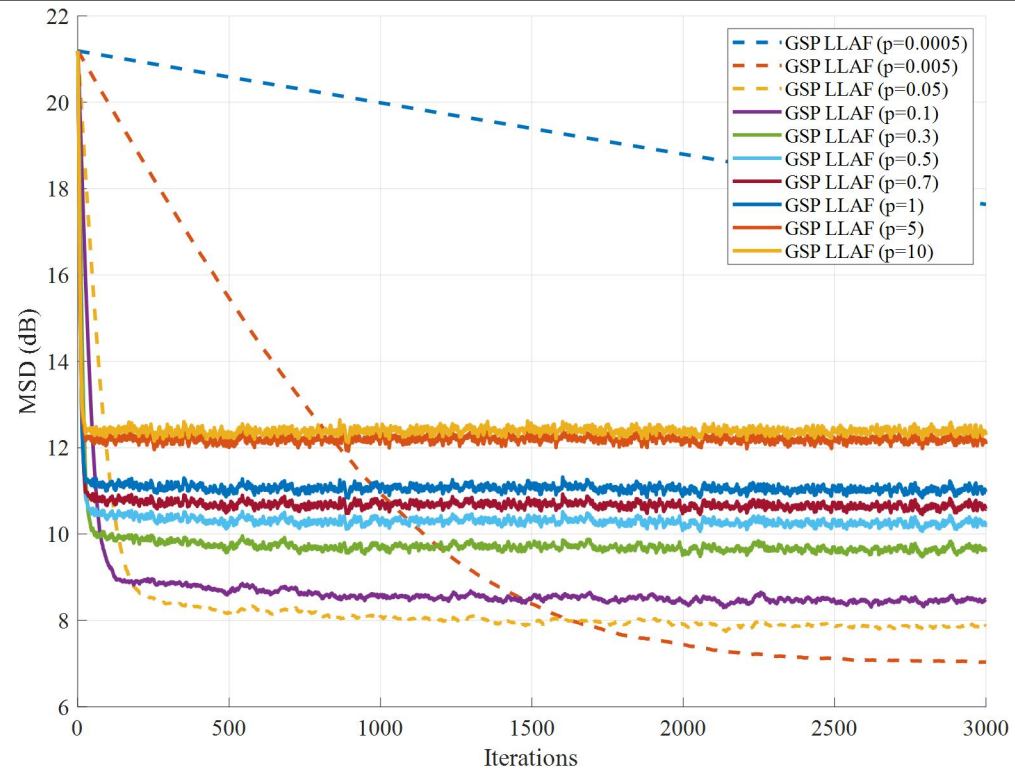

(a)

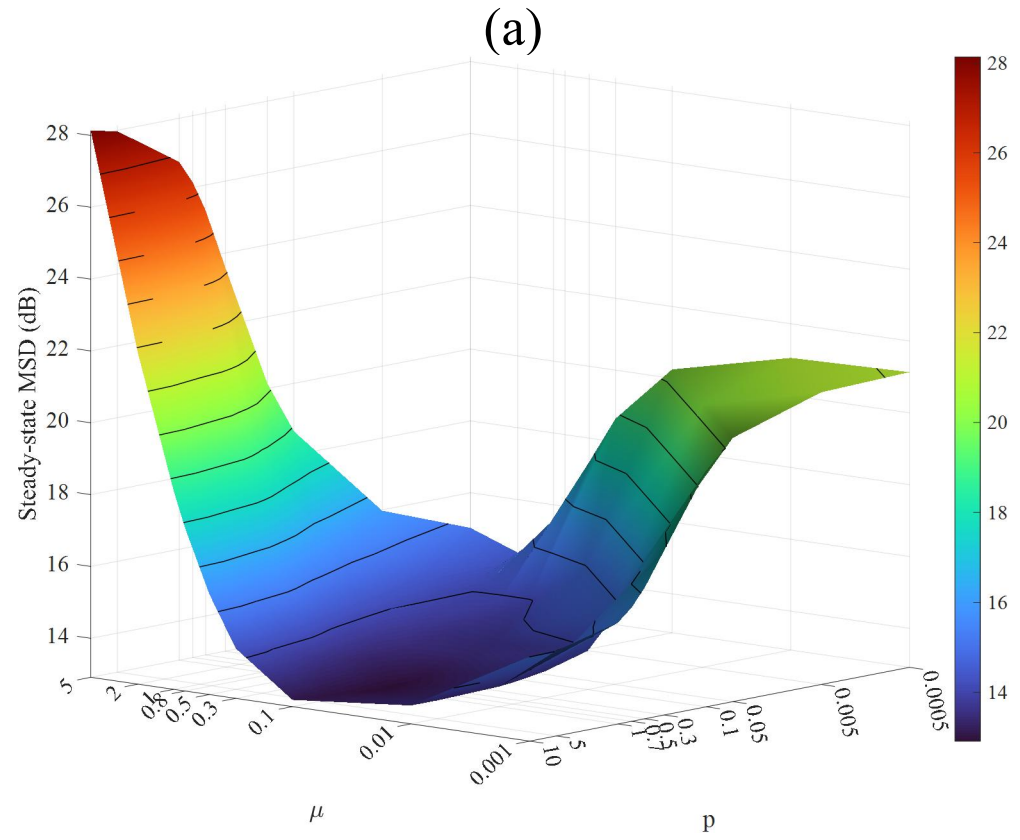

(b)

Fig. 6. Effects of different parameters on the GSP LLAF algorithm (a)Effects of different $p$ values (b) 3D surface of steady-state MSD versus $p$ and $\mu$ .

In practice, $p$ and $\mu$ can be initialized without knowing the ground-truth graph signal. Since $p$ controls the transition of the ln-cosh criterion between its MSE-like and MAE-like behaviors [49], [50], [51], it can be selected according to a robust residual scale. Specifically, over a short initialization window, let [52]

$$\hat{s}_e = \text{median}\left| e - \text{median}(e) \right| + \varepsilon_s \tag{60}$$

and initialize

$$p_0 = \frac{c}{\hat{s}_e} \tag{61}$$

where $c = 1$ is a practical default and $0.5 \le c \le 2$ can be used to adjust the robustness–convergence tradeoff. According to the stability bound and conventional adaptive-filtering guidelines [53], [54], the initial step size can be conservatively selected as

$$\mu_0 = \frac{2\eta}{p_0 \lambda_{\max}(\mathbf{G})} \tag{62}$$

where $\eta = 0.05$ is a conservative default and may be increased when faster tracking is required. A larger $c$ provides stronger suppression of large residuals, whereas a larger $\eta$ accelerates convergence at the cost of increased steady-state fluctuations. These rules require neither the ground-truth graph signal nor prior knowledge of the noise variance and should be regarded as practical initialization guidelines rather than universally optimal values.

*2) GSP MVSS-LLAF Algorithm Parameter Discussion*

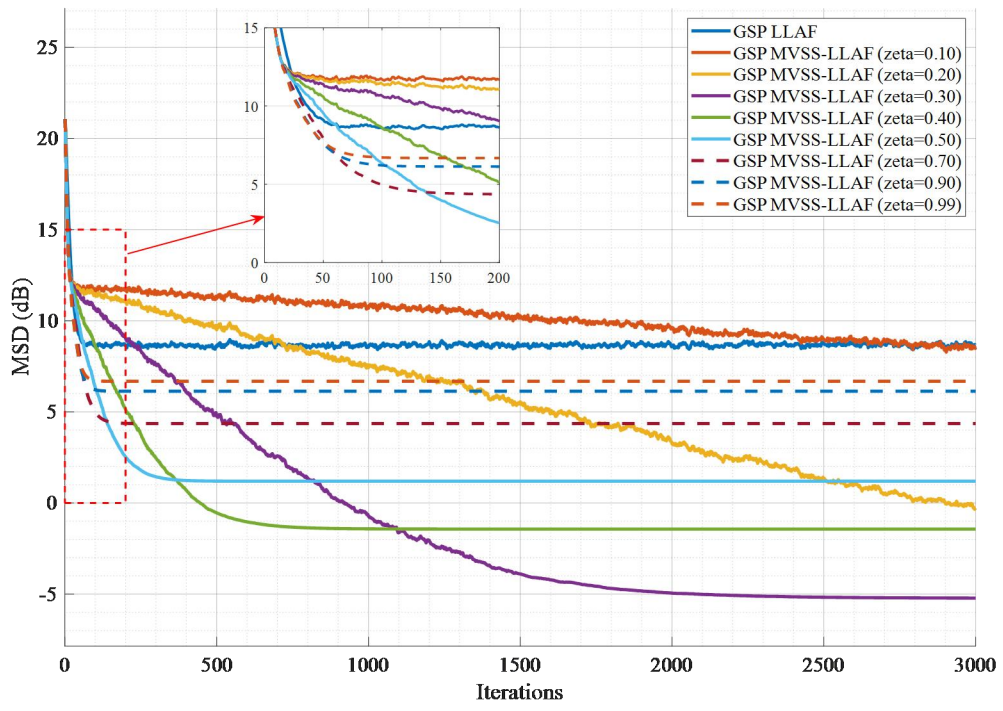

(a)

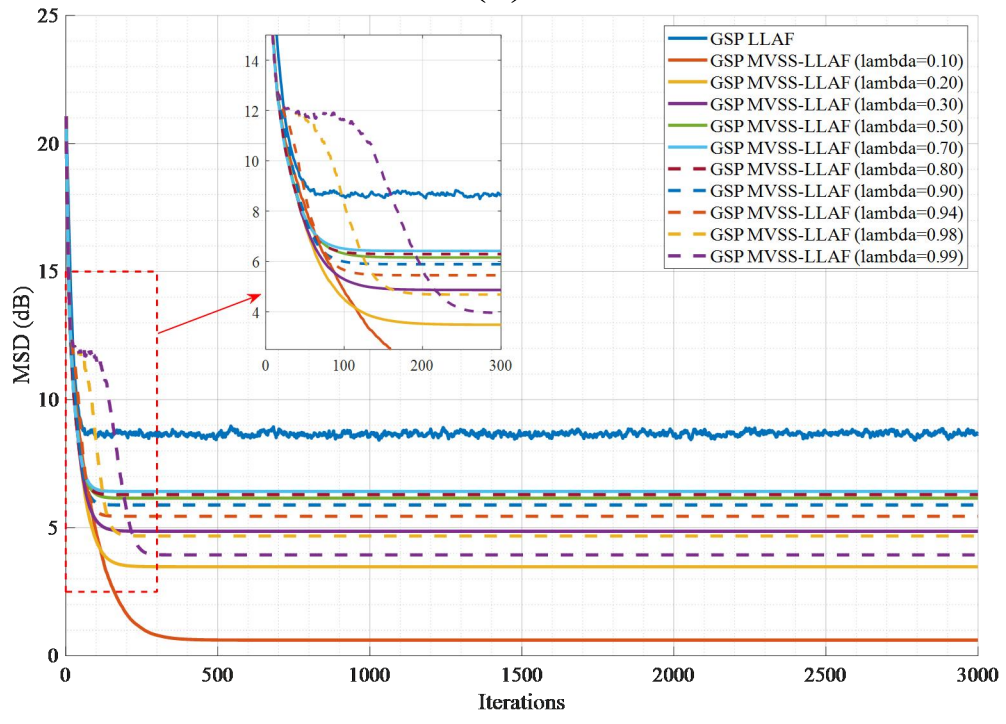

(b)

Fig. 7. The effect of different parameters on the performance of the GSP MVSS-LLAF algorithm (a) Effects of different $\zeta$ values (b) Effects of different $\lambda$ values.

This section investigates the impact of different parameters on the performance of the GSP MVSS-LLAF algorithm through Experiment 2, using the data from $\mathbf{G}_1$. The remaining parameter settings are listed in Table IV. The experiment is conducted under α-stable noise conditions. The simulation results of Experiment 2 are shown in Fig. 7.

TABLE IV
PARAMETER SETTINGS RELATED TO EXPERIMENT 2

| Experiment number | The related parameters | noise environment |
|---|---|---|
| Experiment 2-1 | $\mu_{\text{GSP LLAF}} = 0.3, p = 0.3, \mu_0 = 0.6, \lambda = 0.85$<br>$\zeta \in \{0.10, 0.20, 0.30, 0.40, 0.50, 0.70, 0.90, 0.99\}$ | $\alpha = 1.4$<br>$\gamma = 1$ |
| Experiment 2-2 | $\mu_{\text{GSP LLAF}} = 0.3, p = 0.3, \mu_0 = 0.6, \zeta = 0.9$<br>$\lambda \in \{0.10, 0.20, 0.30, 0.50, 0.70$<br>$, 0.80, 0.90, 0.94, 0.98, 0.99\}$ | $\alpha = 1.4$<br>$\gamma = 1$ |

From Fig. 7(a), the following conclusions can be drawn: as parameter $\zeta$ increases, the convergence speed of the GSP MVSS-LLAF algorithm gradually accelerates, while the steady-state error first decreases and then increases. From Fig. 7(b), the following conclusions can be drawn: as parameter $\lambda$ increases ( $\lambda \le 0.7$ ), the convergence speed of the GSP MVSS-LLAF algorithm gradually accelerates, and the steady-state error gradually rises. However, when $\lambda$ is excessively large ( $\lambda \ge 0.98$ ), an unstable transient state occurs, causing the algorithm's convergence speed to drastically slow down.

*B. Comparison of Algorithm Performance under Different Sample Sizes*

TABLE V
PARAMETER SETTINGS RELATED TO EXPERIMENT 3

| Algorithm | The related parameters | noise environment |
|---|---|---|
| GSP LMS | $\mu = 0.3$ | $\alpha = 1.4$<br>$\gamma = 1$ |
| GSP NLMS | $\mu = 0.07$ | |
| GSP MCC | $\mu = 0.3, \lambda = 0.01$ | |
| GSP GMCC | $\mu = 0.3, \alpha = 1.8, \lambda = 0.01$ | |
| GSP Sign | $\mu = 0.3$ | |
| GSP LLAF | $\mu = 0.3, p = 0.3$ | |
| GSP MVSS-LLAF | $\mu_0 = 0.6, p = 0.3$<br>$\zeta = 0.9, \lambda = 0.85$ | |

As the number of sampling points decreases, it is often observed that the algorithm's convergence speed slows down and the steady-state error increases, which constitutes algorithm performance degradation [14]. Therefore, this section will use Experiment 3 to investigate the performance comparison of the GSP LLAF and GSP MVSS-LLAF algorithms with existing algorithms under different $|\mathbf{S}|$. In Experiment 3, $F = 15$ and $|\mathbf{S}| \in \{18, 20, 22, 25, 30, 40, 50, 60, 70\}$ to generate $\mathbf{G}_2$. The remaining parameter settings are listed in Table V. To ensure fairness in the experiment, all algorithms employ a uniform step size. The experiment is conducted under α-stable noise conditions. The simulation results of Experiment 3 are shown in Fig. 8 and Fig. 9.

Fig. 8 presents a comparison of the algorithm steady-state errors under different $|\mathbf{S}|$, while Fig. 9 shows the algorithm convergence curves for $|\mathbf{S}| = 18$ and $|\mathbf{S}| = 70$. Based on these results, the following conclusions can be drawn: the GSP

LLAF algorithm retains good robustness even when $|\mathbf{S}|$ decreases (under a low $|\mathbf{S}|$ ). In contrast, the performance of the GSP MVSS-LLAF algorithm degrades significantly when $|\mathbf{S}|$ decreases (under a low $|\mathbf{S}|$ ). Specifically, GSP MVSS-LLAF remains convergent at $|\mathbf{S}|=18$ , but its steady-state MSD is higher than that of GSP LLAF; as $|\mathbf{S}|$ increases, its steady-state MSD decreases and becomes lower than that of GSP LLAF when $|\mathbf{S}|>50$ .

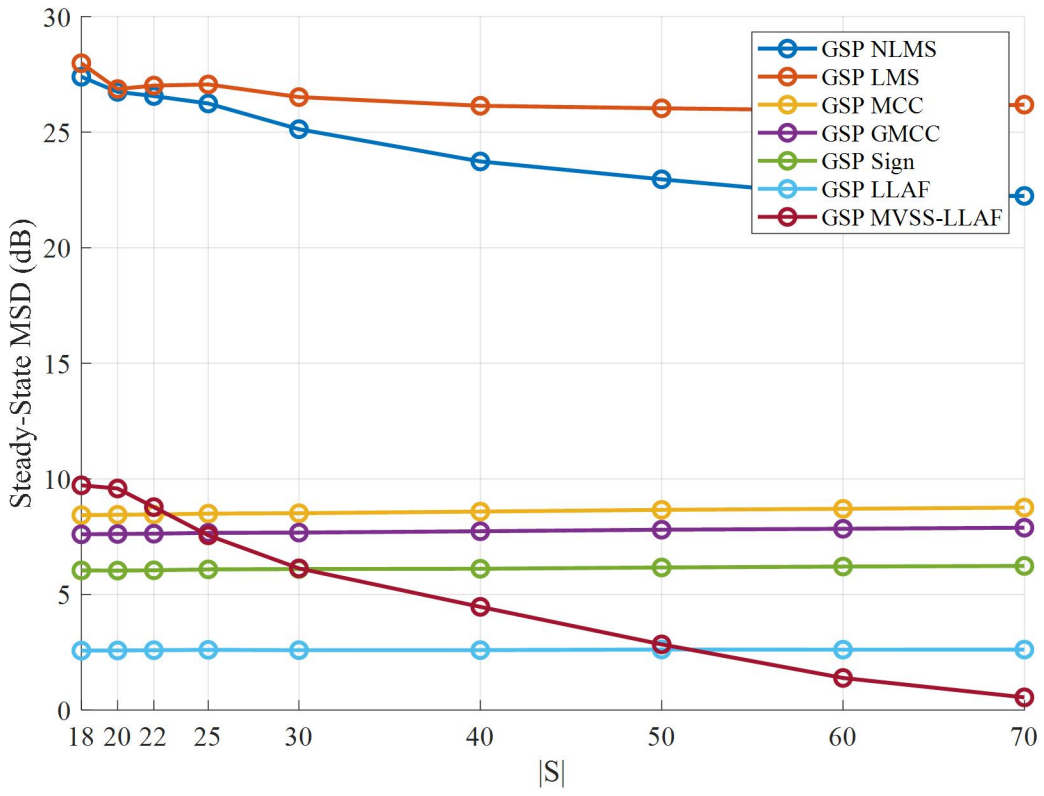

Fig. 8. Steady-state MSD comparison with different $|\mathbf{S}|$ .

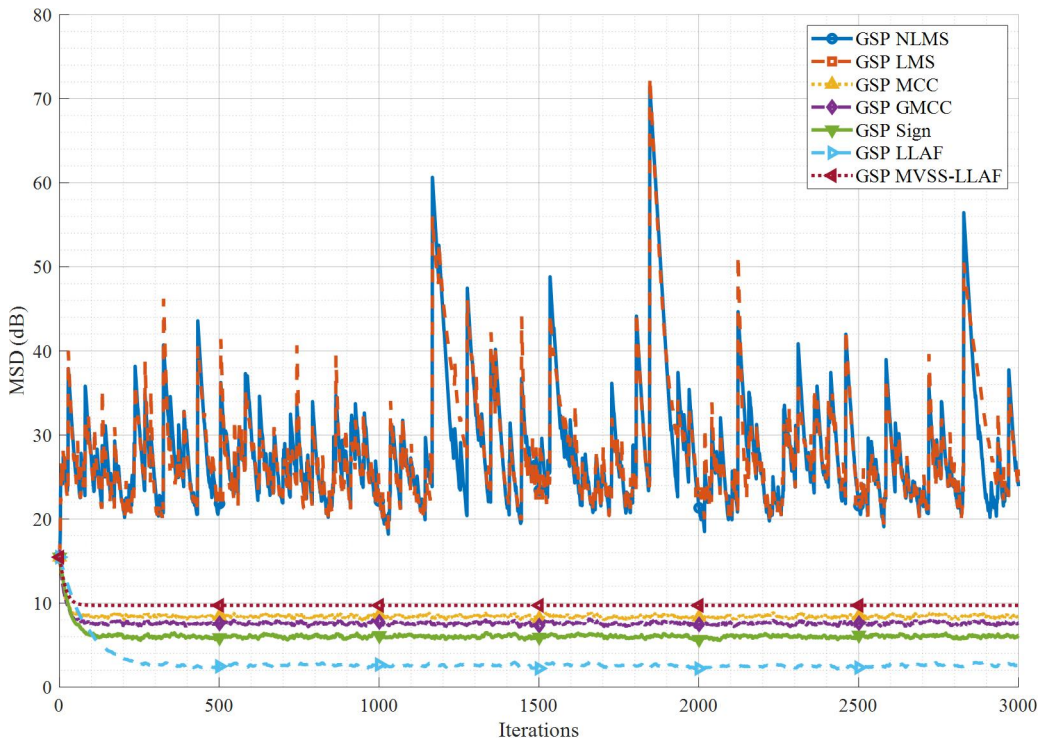

(a)

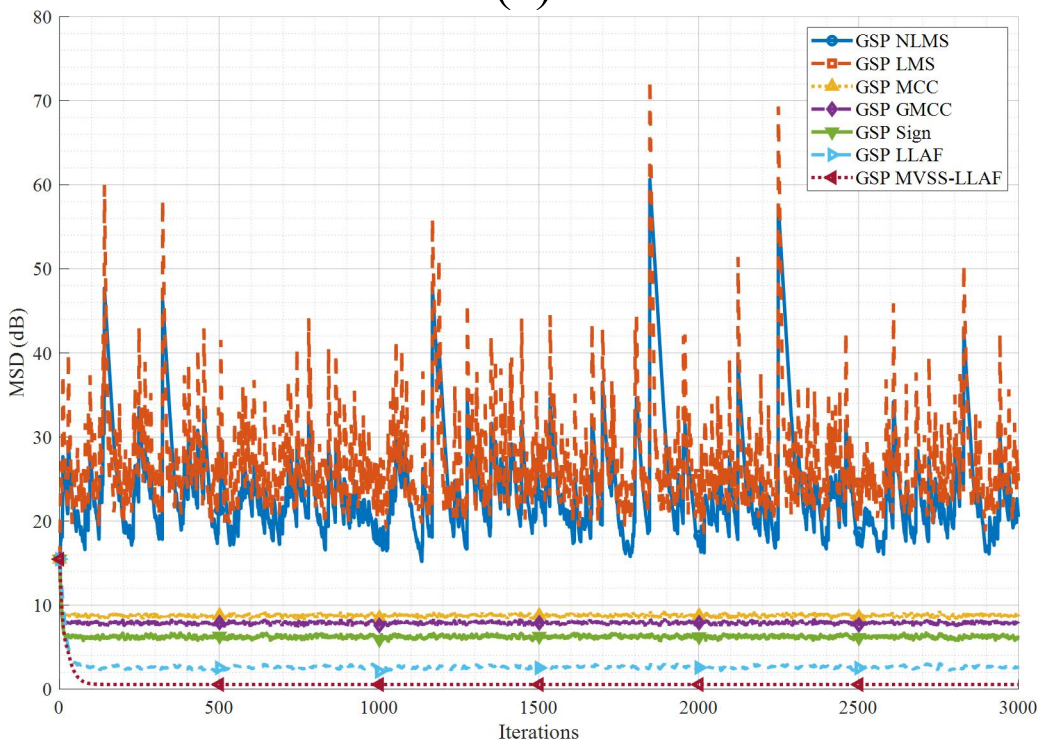

(b)

Fig. 9. Comparison of algorithm convergence curves under different $|\mathbf{S}|$ (a) $|\mathbf{S}|=18$ (b) $|\mathbf{S}|=70$ .

This behavior can be explained from both the graph-sampling and step-size adaptation perspectives. As $|\mathbf{S}|$ decreases, the smallest eigenvalue of $\mathbf{G}=\mathbf{U}_F^T\mathbf{D}_S\mathbf{U}_F$ may approach zero, causing the contraction factor associated with the recursion in (26) to approach unity. Consequently, convergence becomes slower and the finite-iteration estimation error may increase. Moreover, the instantaneous noise-magnitude estimate used in (18), $b[i]=|\mathbf{S}|^{-1}\sum_{k\in\mathbf{S}}|\mathbf{e}_k[i]|$ , is computed from only the sampled nodes. Under the i.i.d. assumption, its variance satisfies $\operatorname{Var}(b[i])=\operatorname{Var}(|\mathbf{e}_k[i]|)/|\mathbf{S}|$. Therefore, reducing $|\mathbf{S}|$ increases the statistical fluctuation of the noise estimate $\beta[i]$, making the variable step-size update less reliable.

*C. Algorithmic Theoretical Performance Verification*

*1) Verification of Step Size Constraint for the GSP LLAF Algorithm*

This section utilizes the data from $\mathbf{G}_3$ to validate the step size constraint range (35). The parameter settings for Experiment 4 are listed in Table VI. The impulse noise environment is simulated using the BG model. The experimental results are shown in Fig. 10.

TABLE VI
PARAMETER SETTINGS RELATED TO EXPERIMENT 4

| The related parameters | noise environment |
|---|---|
| $p=0.3,\mu_{\text{critical}}=7.2897$ | $Pr=0.1$ |
| $k\in\{0.4,0.8,1.0,1.2,1.6\}$ | $\delta_g^2=0.01,\delta_b^2=10^6\delta_g^2$ |

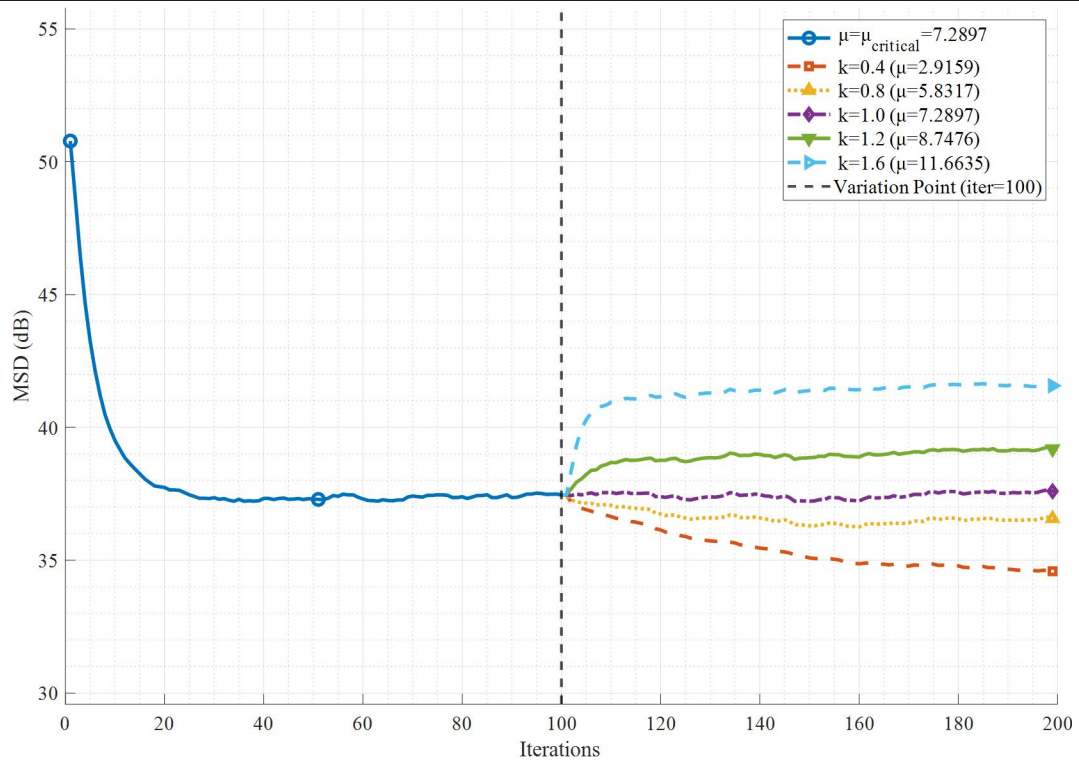

Fig. 10. Verification of the step size constraint range for the GSP LLAF algorithm.

The experiment calculates the critical step size $\mu_{\text{critical}}$ via (35). The GSP LLAF algorithm performs iterative computation with $\mu=\mu_{\text{critical}}$ . After reaching a steady state, at the 100-th iteration, it sets $\mu=k\mu_{\text{critical}}$ , and observes the changes in the algorithm's convergence curve.

By analyzing Fig. 10, the following conclusions can be drawn:

When $k\in\{0.4,0.8\}$ , compared to when $k=1.0$ , the step size of the GSP LLAF algorithm is smaller than the critical value, and it is in a convergent state. When $k\in\{1.2,1.6\}$ , compared to when $k=1.0$ , the step size of the GSP LLAF algorithm is larger than the critical value, and it is in a divergent state—at $k\in\{1.2,1.6\}$, the algorithm's convergence curve continues to rise slowly even after the step size change has stabilized. Therefore, (35) holds.

*2) Verification of Steady-State Error for the GSP LLAF Algorithm*

This section employs the data from $\mathbf{G}_3$ to verify the steady-state error formula (41). The parameter settings for Experiment 5 are listed in Table VII. Since the theoretical derivation is based on the assumption of symmetrically distributed noise (Assumption 3), the experiment is conducted in a Laplace noise environment. The experimental results are shown in Fig. 11.

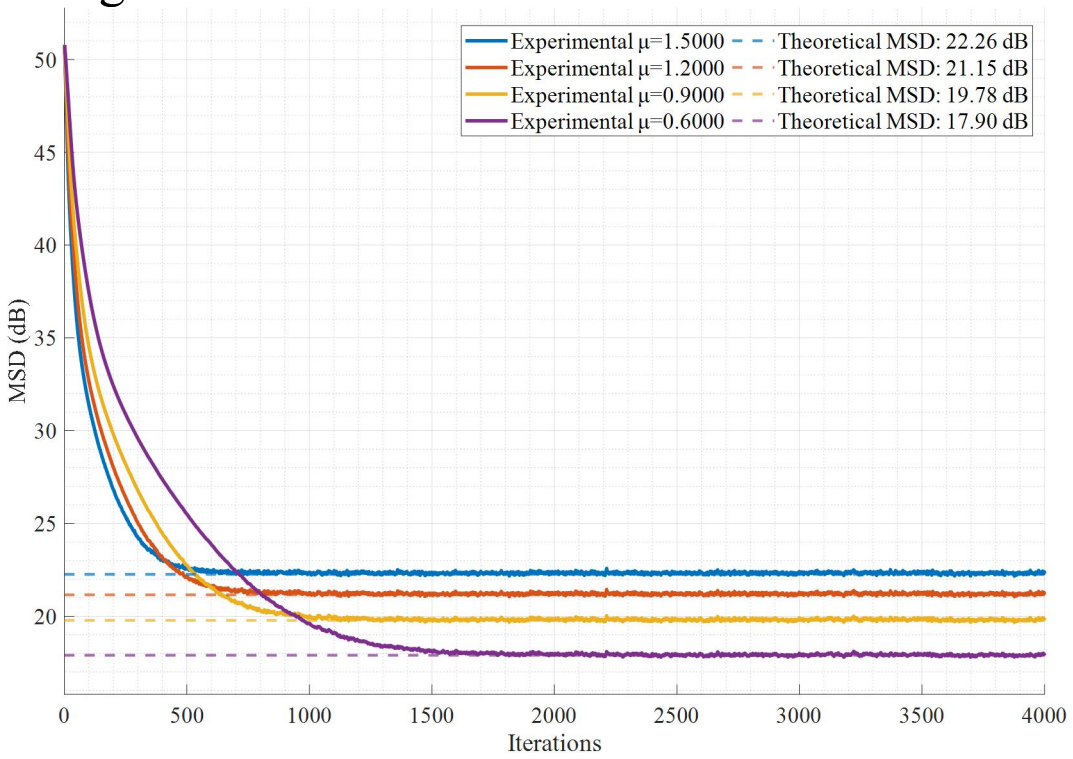

Fig. 11. Verification of theoretical steady-state error for the GSP LLAF algorithm.

TABLE VII

PARAMETER SETTINGS RELATED TO EXPERIMENT 5

| The related parameters | noise environment |
|---|---|
| $p=0.3$<br>$\mu\in\{0.6,0.9,1.2,1.5\}$ | $b_L=1$ |

Using (41), the theoretical steady-state errors for $\mu\in\{0.6,0.9,1.2,1.5\}$ in Experiment 5 are calculated to be 17.90dB, 19.78dB, 21.15dB, and 22.26dB, respectively. In Fig. 11, the algorithm's convergence curves coincide with the theoretical steady-state error baselines upon reaching the steady state, which verifies the validity of (41).

*3) Steady-State Validation via Mean Absolute Deviation*

This section employs the $\mathbf{G}_3$ dataset to validate the steady-state performance of the GSP LLAF algorithm under infinite-variance noise. The parameter settings for Experiment 6 are provided in Table VIII, and the experimental results are presented in Figs. 12 and 13.

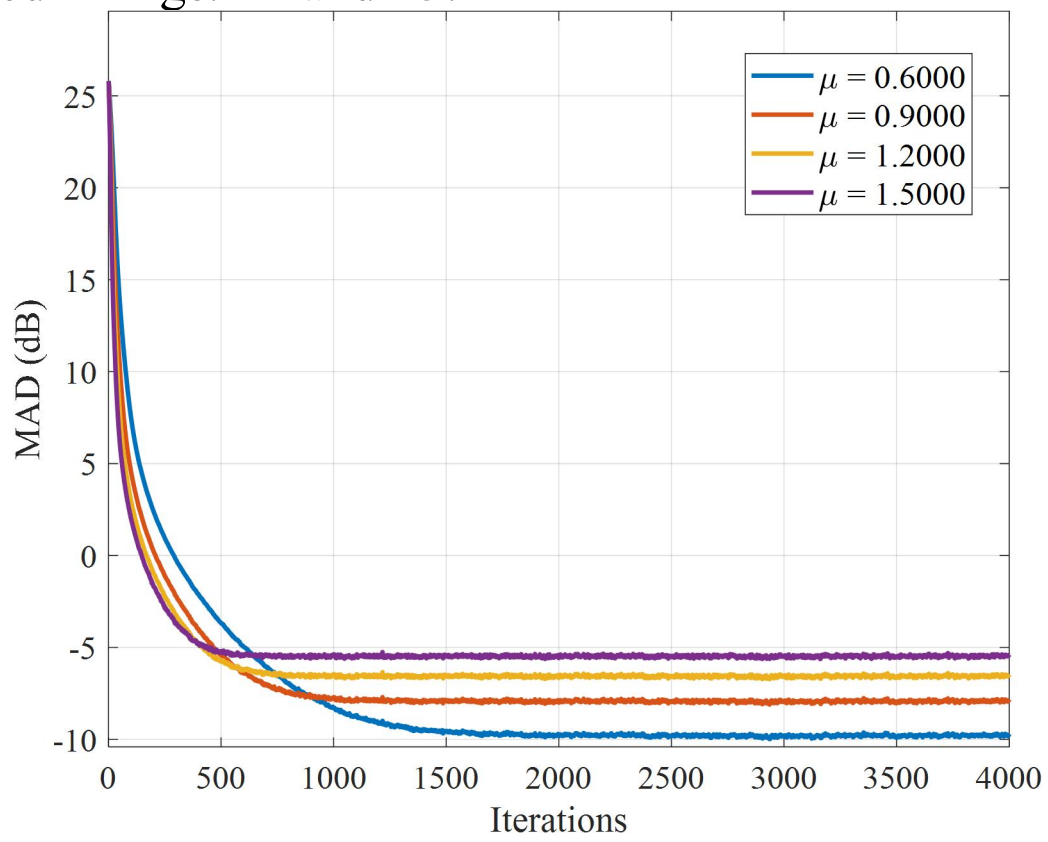

Fig. 12. Steady-state MAD validation of the GSP LLAF algorithm under α-stable noise.

As shown in Figs. 12 and 13, the GSP LLAF algorithm remains capable of converging to steady state even under infinite-variance noise. Furthermore, a smaller μ leads to higher accuracy of the GSP LLAF algorithm, at the cost of an increased number of iterations needed to achieve convergence.

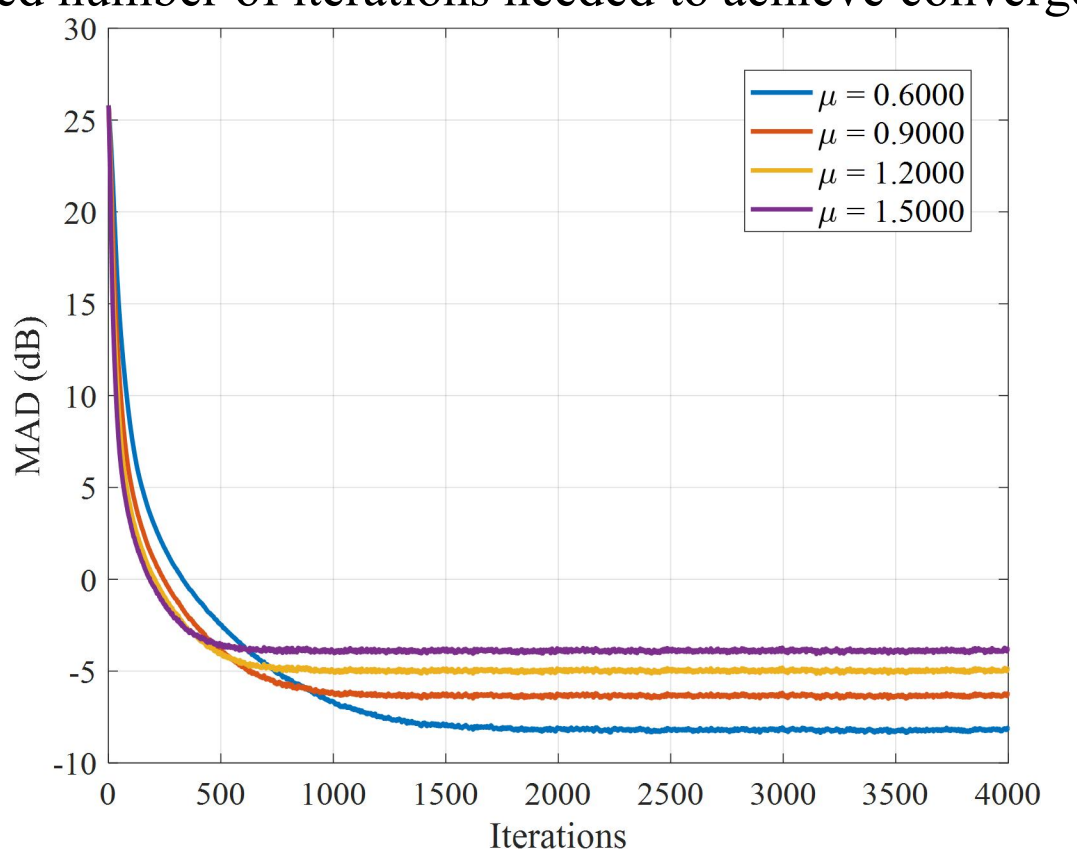

Fig. 13. Steady-state MAD validation of the GSP LLAF algorithm under Cauchy noise.

TABLE VIII

PARAMETER SETTINGS RELATED TO EXPERIMENT 6

| The related parameters | noise environment |
|---|---|
| $p=0.3$<br>$\mu\in\{0.6,0.9,1.2,1.5\}$ | α-stable noise<br>$\alpha=1.4,\gamma=1$<br>Cauchy noise<br>$b_C=1$ |

*D. Static Temperature Graph Signal Estimation Application*

This section employs the data from $\mathbf{G}_3$ to perform static graph signal estimation application under non-Gaussian noise through Experiment 7. Experiment 7 is conducted under both α-stable noise and BG impulse noise conditions. To ensure fairness in the experiment, all algorithms employ a uniform step size, with specific parameter settings listed in Table IX. The experimental results are presented in Fig. 14 and Fig. 15.

TABLE IX

PARAMETER SETTINGS RELATED TO EXPERIMENT 7

| Experiment number | Algorithm | The related parameters | noise environment |
|---|---|---|---|
| Experiment 7-1 | GSP MCC | $\mu=0.98,\lambda=0.01$ | α-stable noise<br>$\alpha=1.4,\gamma=1$ |
| | GSP GMCC | $\mu=0.98,\alpha=1.8,\lambda=0.01$ | |
| | GSP Sign | $\mu=0.98$ | |
| | GSP LLAF | $\mu=0.98, p=0.5$ | |
| | GSP MVSS-LLAF | $\mu_0=0.98, p=0.5$<br>$\zeta=0.8,\lambda=0.9977$ | |
| Experiment 7-2 | GSP MCC | $\mu=0.98,\lambda=0.01$ | BG noise<br>$Pr=0.1$<br>$\delta_g^2=0.01$<br>$\delta_b^2=5\times10^4\delta_g^2$ |
| | GSP GMCC | $\mu=0.98,\alpha=1.8,\lambda=0.01$ | |
| | GSP Sign | $\mu=0.98$ | |
| | GSP LLAF | $\mu=0.98, p=0.5$ | |
| | GSP MVSS-LLAF | $\mu_0=0.98, p=0.5$<br>$\zeta=0.9,\lambda=0.998$ | |

To reflect the distribution and robustness range of the learning curves across multiple independent repeated trials, this section plots quantile-based banded regions at each iteration step $i$, using the empirical 5%–95% quantiles to represent the distribution range covering 90% of the experimental results at that iteration point. Let $M$ denote the number of independent trials, and let $\mathrm{MSD}_t(i)$ represent the mean square error obtained at iteration $i$ in the $t$-th independent trial, where $t=1,\ldots,M$ . For a fixed iteration i, the values $\{\mathrm{MSD}_t(i)\}_{t=1}^{M}$ are arranged in non-descending order as

$$\mathrm{MSD}_1(i)\le \mathrm{MSD}_2(i)\le\cdots\le \mathrm{MSD}_M(i) \tag{63}$$

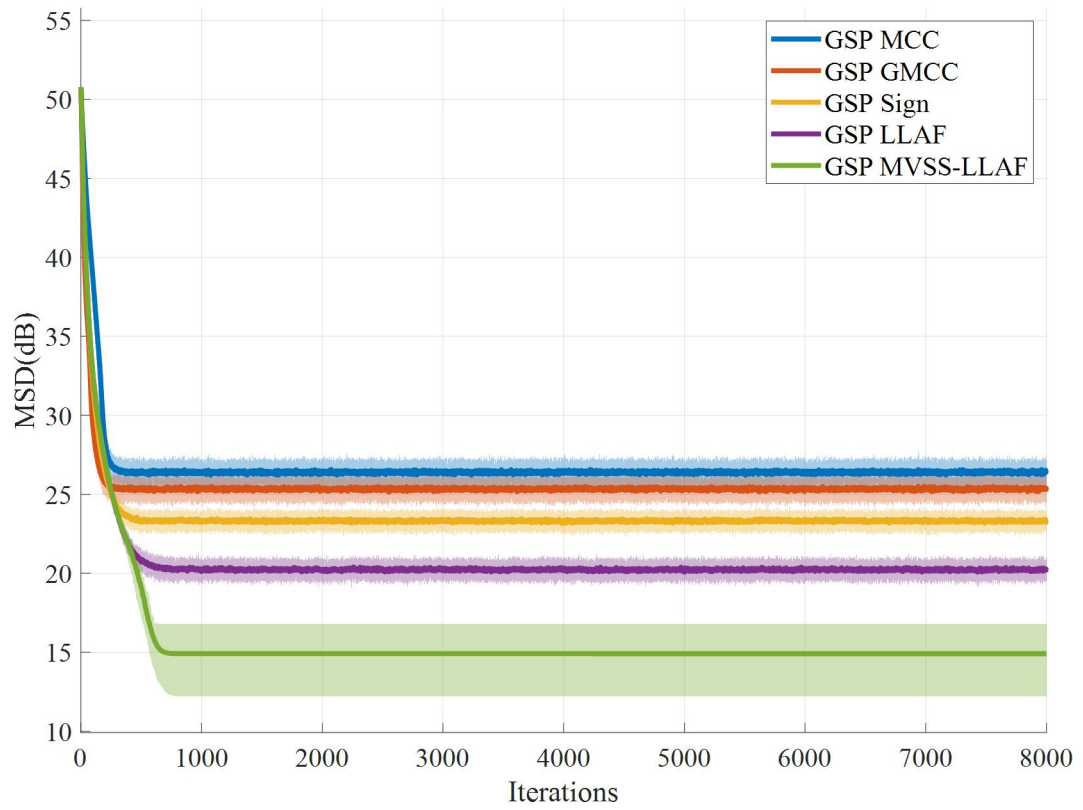

Fig. 14. Algorithm performance comparison under α-stable noise.

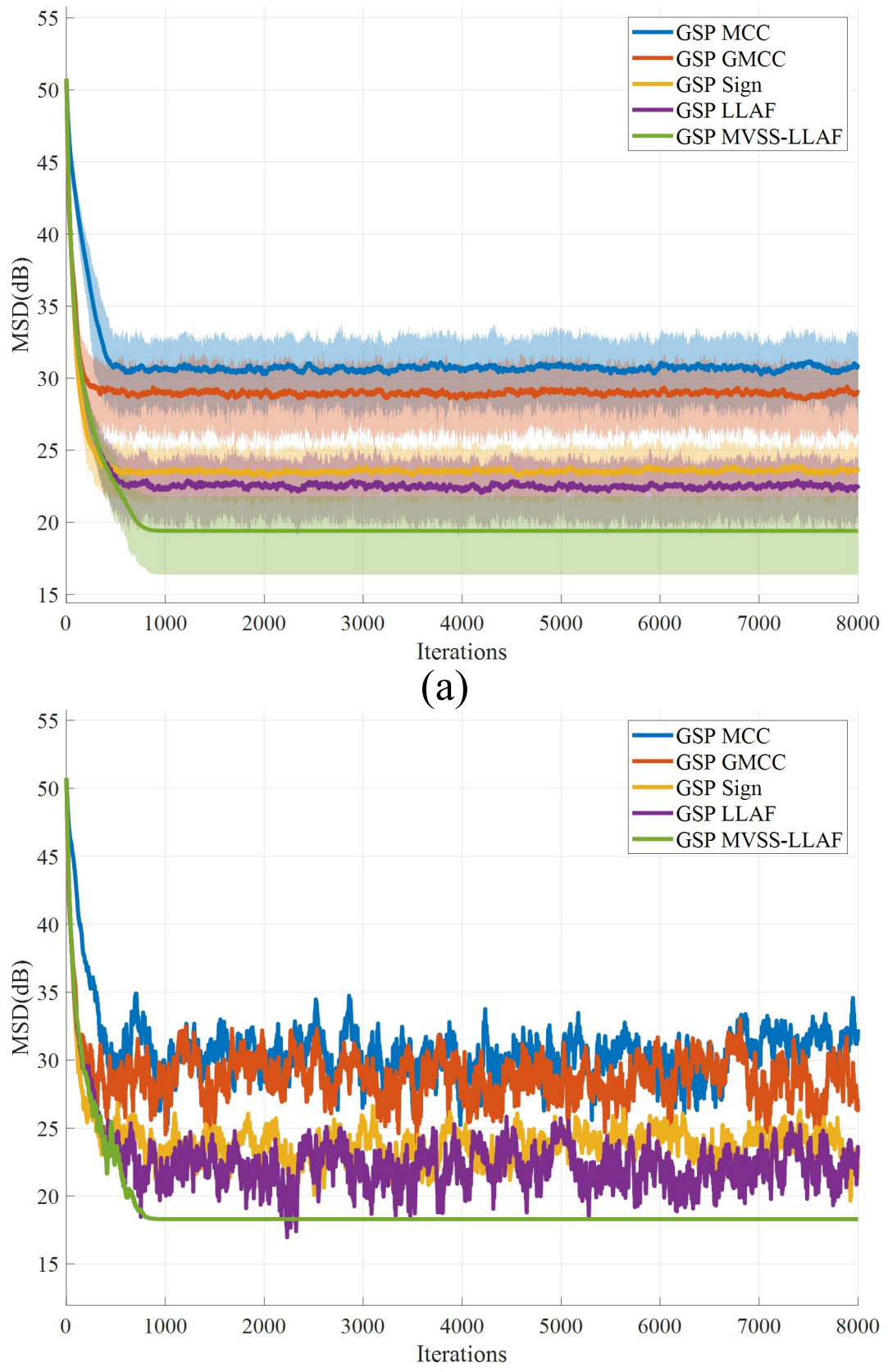

(a)

(b)

Fig. 15. Algorithm performance comparison under BG noise (a) $M=100$ (b) $M=1$ .

The empirical quantile $q_p(i)$ at probability $p$ ( $0<p<1$ ) is taken as the corresponding order statistic

$$q_p(i)=\mathrm{MSD}_{(\lceil pM\rceil)}(i) \tag{64}$$

where $\lceil\cdot\rceil$ denotes the ceiling function. This paper employs $p=0.05$ and $p=0.95$ , obtaining the lower bound $q_{0.05}(i)$ and upper bound $q_{0.95}(i)$ , respectively. The boundary curves of the shaded band are defined as

$$\begin{cases}\mathrm{upper}_{\mathrm{dB}}=10\log_{10}\left(q_{0.95}(i)\right)\\ \mathrm{lower}_{\mathrm{dB}}=10\log_{10}\left(q_{0.05}(i)\right)\end{cases} \tag{65}$$

Fig. 15(a) and Fig. 14 present the results from 100 independent repeated trials under BG noise and α-stable noise, respectively. The following conclusions can be drawn: the performance of the GSP MCC algorithm degrades drastically under BG noise. Moreover, both the GSP LLAF and GSP MVSS-LLAF algorithms exhibit superior performance. Furthermore, this section utilizes (65) to examine the empirical quantile bands of the learning curves for each algorithm in the experiment, thereby strengthening the persuasiveness of the conclusions. Fig. 15(b) shows the results from a single independent trial under BG noise. It can be observed that upon reaching the steady state, the step size of the GSP MVSS-LLAF algorithm converges to a minimal value, which can be interpreted as the algorithm halting its updates, effectively avoiding oscillations during the steady state.

### *E. Dynamic Temperature Graph Signal Estimation Application*

This section employs the data from $\mathbf{G}_4$ to perform dynamic graph signal estimation under non-Gaussian noise environments through Experiment 8, where each hour is treated as one iteration, and the algorithms output an online estimate. Experiment 8 is conducted under α-stable noise, BG impulse noise, Laplace noise, and Cauchy noise conditions. To ensure fairness in the experiment, all algorithms employ a uniform step size, with specific parameter settings listed in Table X. Fig. 16, Fig. 17, Fig. 18, and Fig. 19 show the temperature estimation results for one of the nodes. The memory usage of the two algorithms during a single iteration is shown in Table XI.

TABLE X
PARAMETER SETTINGS RELATED TO EXPERIMENT 8

| Algorithm | The related parameters | noise environment |
|---|---|---|
| GSP Sign | $\mu=3$ | α-stable noise<br>$\alpha=1.4,\gamma=0.2$ |
| | | BG noise<br>$Pr=0.1$<br>$\delta_g^2=0.01,\delta_b^2=5\times10^4\delta_g^2$ |
| GSP LLAF | $\mu=3,p=0.3$ | Cauchy noise<br>$b_C=0.2$ |
| | | Laplace noise<br>$b_L=0.2$ |

From Fig. 16 to Fig. 19, it can be concluded that the GSP LLAF algorithm is capable of accurately and efficiently tracking and estimating time-varying graph signals across various non-Gaussian noise environments. Compared to the GSP Sign algorithm, the GSP LLAF algorithm demonstrates higher estimation precision when errors are small. In this experiment, this manifests specifically as follows: in regions where the graph signal curve changes relatively slowly, the estimated curve of the GSP LLAF algorithm aligns more closely with the target curve. In contrast, the GSP Sign algorithm exhibits excessive update steps at the peaks and oscillatory behavior at the troughs.

TABLE XI
COMPARISON OF ALGORITHM MEMORY USAGE

| Algorithm | Memory usage |
|---|---|
| GSP Sign | 335.9766 (KB) |
| GSP LLAF | 335.9844 (KB) |

From Table XI, it can be observed that the memory usage of the GSP LLAF algorithm in Experiment 8 is only about 0.0023% higher than that of the GSP Sign algorithm. Combined with the computational complexity analysis in Section IV, we therefore consider the additional cost incurred by the GSP LLAF algorithm to be acceptable.

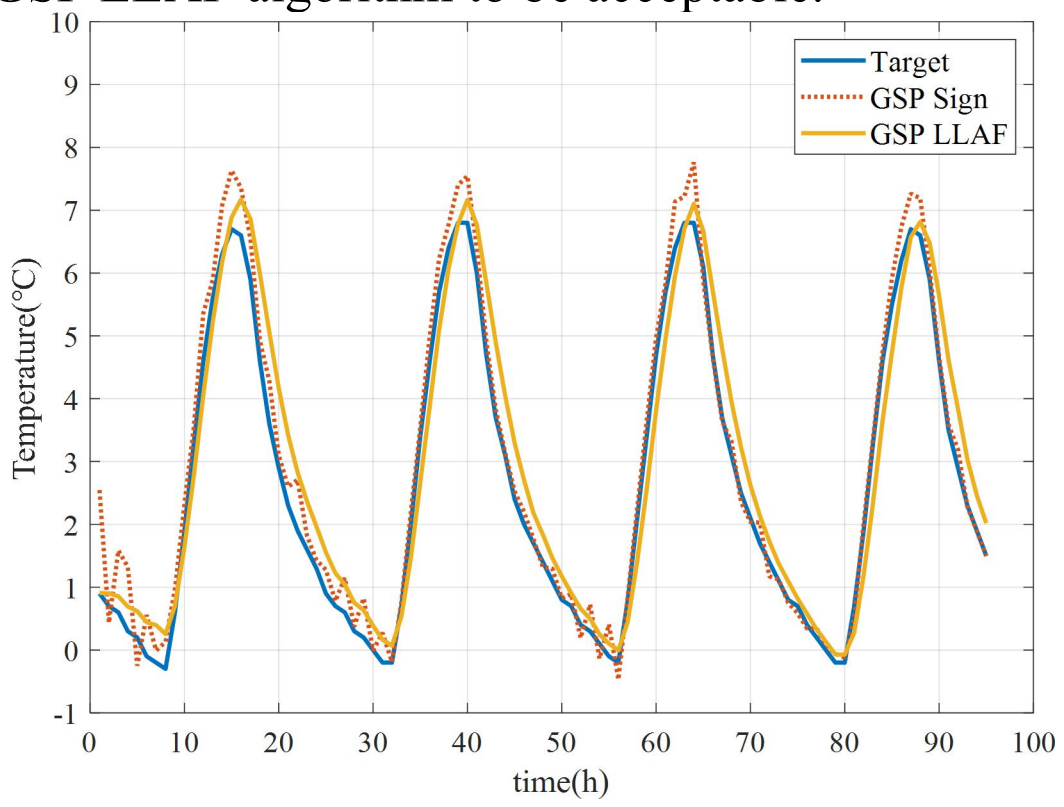

Fig. 16. Algorithm comparison under α-stable noise.

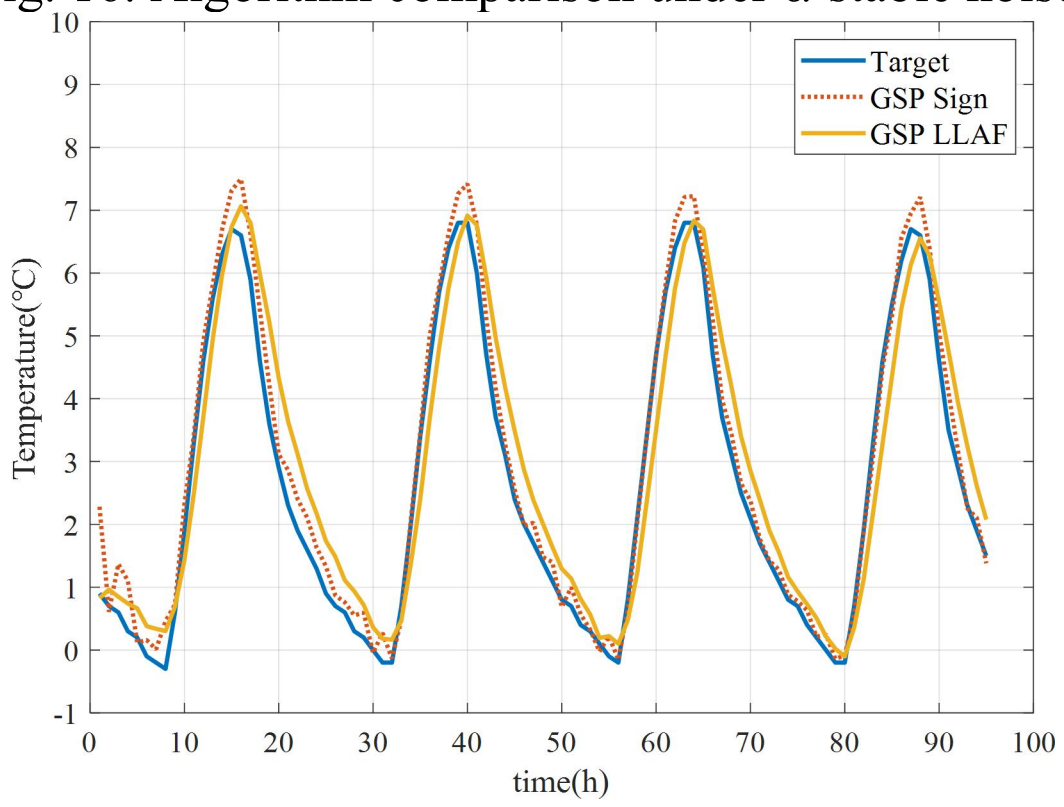

Fig. 17. Algorithm comparison under BG noise.

## VI. CONCLUSION

This paper first proposes an offline/online graph signal estimation algorithm, GSP LLAF, and its variable step-size variant, GSP MVSS-LLAF, designed for environments with non-Gaussian noise and limited observation points. Subsequently, experimental discussions on algorithm parameters, theoretical analysis, and experimental validation were conducted for both the GSP LLAF and GSP MVSS-LLAF algorithms. Finally, the practical performance of the algorithms in offline/online graph signal estimation applications under non-Gaussian noise was compared using randomly generated signals and publicly available temperature datasets. Experimental results demonstrate that the GSP LLAF and GSP MVSS-LLAF algorithms offer advantages such as low complexity, high steady-state accuracy, and robustness.

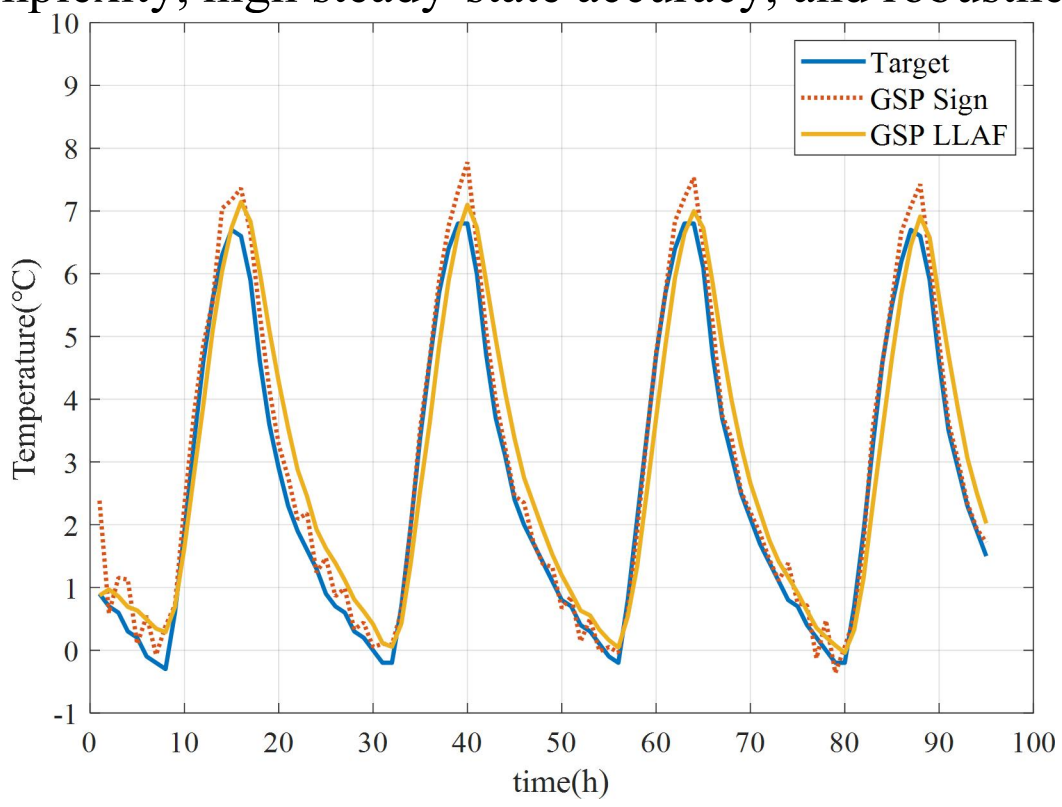

Fig. 18. Algorithm comparison under Cauchy noise.

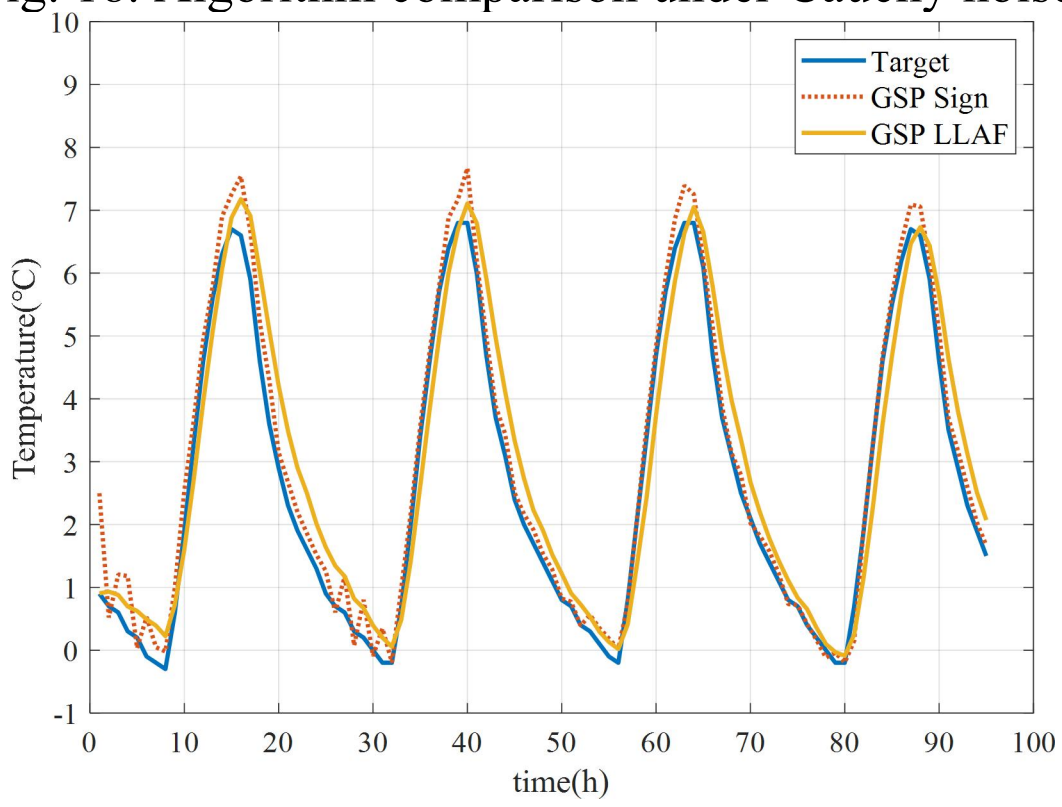

Fig. 19. Algorithm comparison under Laplace noise.

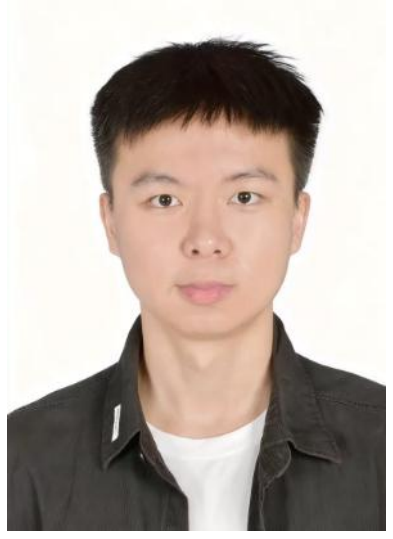

**Chong Zhang** received the Bachelor of Engineering degree in Electrical Engineering and Automation from Hebei University of Technology, Tianjin, China, in 2024. He is currently pursuing the Master of Engineering degree in Electrical Engineering at the School of Electrical Engineering, Southwest Jiaotong University, Chengdu, China. His current research interests include graph signal processing and adaptive filtering algorithms.

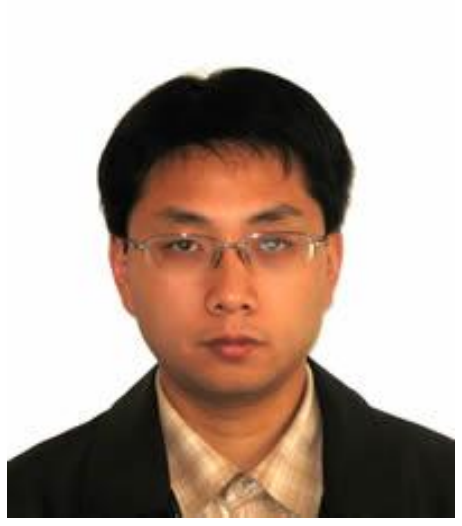

**Haiquan Zhao** (Senior Member, IEEE) received the B.S. degree in applied mathematics and the M.S. and Ph.D. degrees in signal and information processing from Southwest Jiaotong University, Chengdu, China, in 1998, 2005, and 2011, respectively. Since 2012, he has been a Professor with the School of Electrical Engineering, Southwest Jiaotong University. From 2015 to 2016, he was a Visiting Scholar with the University of Florida, Gainesville, FL, USA. He is the author or coauthor of more than 280 international journal papers (SCI indexed) and owns 56 invention patents. His current research interests include information theoretical learning, adaptive filters, adaptive networks, active noise control, Kalman filters, machine learning, and artificial intelligence.

Dr. Zhao has won several provincial and ministerial awards and many best paper awards at international conferences or IEEE TRANSACTIONS. He has served as an Active Reviewer for several IEEE TRANSACTIONS, IET series, signal processing, and other international journals. He is currently a Handling Editor of Signal Processing, and also an Associate Editor for IEEE Transaction on Audio, Speech and Language Processing, IEEE TRANSACTIONS ON SYSTEMS, MAN AND CYBERNETICS: SYSTEM, IEEE SIGNAL PROCESSING LETTERS, IEEE SENSORS JOURNAL, and IEEE OPEN JOURNAL OF SIGNAL PROCESSING.